\documentclass[preprint2]{aastex631}

\usepackage{bm} 
\usepackage{amsmath}
\usepackage{multirow}
\usepackage{ulem}
\usepackage{float, placeins, xcolor}
\usepackage[version=4]{mhchem}

\newcommand{\Secref}[1]{\hyperref[#1]{Section~\ref*{#1}}}
\newcommand{\Appref}[1]{\hyperref[#1]{Appendix~\ref*{#1}}}
\newcommand{\breads}{\texttt{breads} }
\newcommand{\breadsnospace}{\texttt{breads}}

\shorttitle{Improving Detection Sensitivities with Moderate-Resolution IFS}
\shortauthors{Agrawal et al.}

\graphicspath{{./}{figs/}{pdf/}{eps/}{figures/}}

\received{December 10, 2022}
\revised{April 19, 2023; May 11, 2023}
\accepted{May 12, 2023}

\submitjournal{AJ}

\newcommand{\reviewedit}[1]{#1}
\newcommand{\secondreviewedit}[1]{#1}

\begin{document}

\title{Detecting Exoplanets Closer to Stars with Moderate Spectral Resolution Integral-Field Spectroscopy}

\newcommand{\caltech}{Department of Astronomy, California Institute of Technology, Pasadena, CA 91125, USA}
\newcommand{\jpl}{Jet Propulsion Laboratory, California Institute of Technology, 4800 Oak Grove Dr., Pasadena, CA 91109, USA}
\newcommand{\ucsd}{Department of Physics \& Center for Astrophysics and Space Sciences, University of California, San Diego, La Jolla, CA 92093, USA}

\correspondingauthor{Shubh Agrawal}
\email{shubh@sas.upenn.edu}

\author[0000-0003-2429-5811]{Shubh Agrawal}
\affiliation{\caltech}
\affiliation{Department of Physics and Astronomy, University of Pennsylvania, Philadelphia, PA 19104, USA}

\author[0000-0003-2233-4821]{Jean-Baptiste Ruffio}
\affiliation{\caltech}
\affiliation{\ucsd}

\author[0000-0002-9936-6285]{Quinn M. Konopacky}
\affiliation{\ucsd}

\author[0000-0003-1212-7538]{Bruce Macintosh}
\affiliation{Kavli Institute for Particle Astrophysics and Cosmology, Stanford University, Stanford, CA 94305, USA}
\affiliation{University of California Observatories, University of California Santa Cruz, Santa Cruz, CA 95064, USA}

\author[0000-0002-8895-4735]{Dimitri Mawet}
\affiliation{\caltech}
\affiliation{\jpl}

\author[0000-0001-6975-9056]{Eric L. Nielsen}
\affiliation{Department of Astronomy, New Mexico State University, P.O. Box 30001, MSC 4500, Las Cruces, NM 88003, USA}

\author[0000-0002-9803-8255]{Kielan K. W. Hoch}
\affiliation{\ucsd}
\affiliation{Space Telescope Science Institute, 3700 San Martin Drive, Baltimore, MD 21218, USA}

\author[0000-0003-2232-7664]{Michael C. Liu}
\affiliation{Institute of Astronomy, University of Hawaii, 2860 Woodlawn Drive, Honolulu, HI 96822, USA}

\author[0000-0002-7129-3002]{Travis S. Barman}
\affiliation{Lunar and Planetary Laboratory, University of Arizona, Tucson, AZ 85721, USA}

\author[0000-0001-5684-4593]{William Thompson}
\affiliation{University of Victoria,
    Department of Physics and Astronomy,
    3800 Finnerty Rd,
    Victoria, BC V8P 5C2, Canada}
    
\author[0000-0002-7162-8036]{Alexandra Z. Greenbaum}
\affiliation{IPAC, Mail Code 100-22, California Institute of Technology, 1200 E. California Blvd., Pasadena, CA 91125, USA}

\author[0000-0002-4164-4182]{Christian Marois}
\affiliation{National Research Council of Canada Herzberg, 5071 West Saanich Rd, Victoria, BC, V9E 2E7, Canada}
\affiliation{University of Victoria, 3800 Finnerty Rd, Victoria, BC, V8P 5C2, Canada}

\author[0000-0001-9004-803X]{Jenny Patience}
\affiliation{School of Earth and Space Exploration, Arizona State University, Tempe, AZ 85282, USA}

\begin{abstract}

While radial velocity surveys have demonstrated that the population of gas giants peaks around $3~\text{au}$, the most recent high-contrast imaging surveys have only been sensitive to planets beyond \reviewedit{$\sim$} $10~\text{au}$. Sensitivity at small angular separations from stars is currently limited by the variability of the point spread function.
We demonstrate how moderate-resolution integral field spectrographs can detect planets at smaller separations (\reviewedit{$\lesssim$} $0.3$ arcseconds) by detecting the distinct spectral signature of planets compared to the host star.
Using OSIRIS ($R$ \reviewedit{$\approx$} 4000) at the W.M. Keck Observatory, we present the results of a planet search via this methodology around \reviewedit{20} young targets in the Ophiuchus and Taurus star-forming regions. We show that OSIRIS can outperform high-contrast coronagraphic instruments equipped with extreme adaptive optics and non-redundant masking in the $0.05-0.3$ arcsecond regime.
As a proof of concept, we present the $34\sigma$ detection of a high-contrast M dwarf companion at $\approx0.1$" with flux ratio \reviewedit{of} $\approx0.92\%$ around \reviewedit{the field F2 star} HD 148352.
We developed an open-source Python package, \breadsnospace, for the analysis of moderate-resolution integral field spectroscopy data in which the planet and the host star signal are jointly modeled. The diffracted starlight continuum is forward-modeled using a spline model, which removes the need for prior high-pass filtering or continuum normalization. The code allows for analytic marginalization of linear hyperparameters, simplifying posterior sampling of other parameters (e.g., radial velocity, effective temperature).
This technique could prove very powerful when applied to integral field spectrographs like NIRSpec on the JWST and other \reviewedit{upcoming} first light instruments on the future Extremely Large Telescopes.

\end{abstract}

\keywords{\href{http://astrothesaurus.org/uat/387}{Direct Imaging (387)}; \href{http://astrothesaurus.org/uat/489}{Exoplanet detection methods (489)}; \href{http://astrothesaurus.org/uat/2369}{High contrast techniques (2369)}; \href{http://astrothesaurus.org/uat/2096}{High resolution spectroscopy (2096)}}

\section{Introduction}
\label{sec:introduction}

Using the combination of adaptive optics, coronagraphy, and image processing, large direct imaging surveys have detected a dozen of exoplanets at semi-major axes \secondreviewedit{$\gtrsim$} 10 au \citep[e.g.,][]{bowler_imaging_2016, nielsen_gemini_2019, Vigan2021}. These surveys have made use of imagers or low-resolution spectrographs \citep[e.g.,][]{macintosh_first_2014} combined with various speckle subtraction algorithms \citep[e.g.,][]{Cantalloube2020}. The most common observing strategies are angular and/or spectral differential imaging \citep[ADI or SDI;][]{marois_efficient_2000, Lui2004, marois_angular_2006}, which rely on the different behavior between the speckle noise and the companion signal such as the chromatic magnification of the point spread function or the rotation of the planet signal in pupil tracking mode. Both techniques become less effective at smaller separations where the self-subtraction of the companion signal will dominate. \reviewedit{Reference-star differential imaging} (RDI) has been shown to perform better at smaller separations \citep{Xuan2018, Wahhaj2021}, but it often remains speckle noise-limited due to the variability of the point spread function (PSF) between the science target and the reference stars. This explains why the exoplanet sensitivity of high-contrast instruments increases sharply around $0.2^{\prime\prime}$. Additionally, the physical size of a coronagraphic mask can prevent the detection of planets at the smallest inner working angles. As an alternative to coronagraphs, non-redundant aperture masking \citep[e.g.,][]{Kraus2008, Tuthill_2000, Wallace2020} can detect companions at very small separations but is usually limited to brighter companions. 

Direct imaging surveys are already benefiting from the synergies with indirect detection methods such as long-period radial velocity (RV) \citep{Mawet2019, Llop-Sayson2021} and astrometric monitoring from Gaia \citep{DeRosa2019, Currie2020}. Because indirect methods can place priors on the existence of planets and their location, such synergies will allow us to trade large fields of view (FOVs) and low-resolution spectrographs for smaller FOVs with higher spectral resolution. Recent results from RV surveys have shown that the population of Jupiters peaks around 3 au, explaining the relative paucity of directly detected planets \citep{Fernandes2019,fulton_california_2021}. Direct imaging surveys have also been showing an increased frequency of planets toward smaller semi-major axis ($5-10$ au) highlighting the need to improve the sensitivity at smaller projected separations \citep{nielsen_gemini_2019, Vigan2021}. 

Moderate to high-resolution spectroscopy has often been used for atmospheric studies of known exoplanets. Some of the pioneering works include \citet{Konopacky_2013}, who resolved water and carbon monoxide spectral lines for the first time in the atmosphere of HR 8799 c with \reviewedit{Keck/OSIRIS} ($R\approx4,000$). Then, \citet{Snellen2014} for example measured the spin of $\beta$ Pictoris b using \reviewedit{VLT/CRIRES} ($R\approx100,000$). 
The advantage of integral field spectrographs (IFS) is that they can also be used for planet detection using similar techniques. 
When equipped with moderate-resolution spectroscopy, an IFS can leverage the distinct spectral signatures of exoplanet atmospheres compared to the blended spectra of their host stars.
Using cross-correlation techniques, moderate to high-resolution spectroscopy (i.e., $R$ \secondreviewedit{$\gtrsim$} 3,000) is not sensitive to the continuum variability of the speckle noise (i.e., the diffracted starlight from the host star) which is the limiting factor for high-contrast instruments at small projected separations. 
This approach promises to be a game changer in the era of new integral field spectrographs on Extremely Large Telescopes \citep[ELTs;][]{Houlle2021} and onboard JWST which is sensitive to cooler and older planets \citep{Llop-Sayson2021, Patapis2022}. For example, 
\citet{Llop-Sayson2021} simulated that JWST/NIRSpec ($R\sim2,700$), a moderate resolution IFS, will allow the first direct detections of true mature cool Jupiters previously detected from RV searches. These future facilities, including JWST and first-light instruments on the ELTs, are poised to become the next generation of planet-detection facilities.

As a demonstration, we have been conducting a pathfinder survey of the Ophiuchus and Taurus star-forming regions using Keck/OSIRIS. In this work, we present a mid-course sensitivity analysis of \reviewedit{20} stars to demonstrate the improved performance of moderate spectral resolution IFSs. The planet population analysis is left to be discussed in future work.
We are also introducing an innovative open-source package (\breadsnospace\footnote{\url{https://github.com/jruffio/breads}}\footnote{\url{https://github.com/shubhagrawal30/using-breads}}) for moderate to high-resolution spectroscopy building on earlier work \citep{Ruffio2019, Ruffio2021}. This package uses a forward modeling approach to cross-correlation and includes a built-in analytical marginalization of linear parameters.

The observations and pre-reduction steps are described in Section \ref{sec:observations}. Then, the data analysis and sensitivity calculation are presented in Section \ref{sec:datareduction} and the results are shown in Section \ref{sec:results}. Finally, we discuss the potential of this approach in Section \ref{sec:discussion}.

\section{Observations}
\label{sec:observations}

\subsection{Target selection}

At distances of 130-140 pc and ages of 1-2 Myr, Taurus \citep[][]{kenyon_pre-main-sequence_1995} and \reviewedit{Ophiuchus \citep[][]{Wilking2005}} are the youngest and closest star-forming regions, making them ideal targets for high-contrast imaging. Indeed, young planets are still hot, and therefore bright in the infrared, before they cool down with time. For example, a $10 M_\mathrm{Jup}$ companion at 10 au would have an apparent magnitude of $\sim15$ \reviewedit{in K band} and projected separation of $\sim70\,\mathrm{mas}$ \reviewedit{assuming a BT-Settl evolutionary grid \citep{allard_bt-settl_2014} with an age of 2 Myr at 140 pc}. It would therefore be detectable with relatively high flux ratios ($\sim10^{-3}$) for stars with apparent K-band magnitude around 8, which is achievable with OSIRIS.
\reviewedit{In general, we chose the brightest (by $K_\text{mag}$) targets with a visible magnitude $R_\text{mag}<14$} to ensure good adaptive optics correction with a natural guide star. We then selected the more massive stars, because they are more likely to host a gas giant planet according to existing direct imaging surveys \citep{nielsen_gemini_2019}.
\reviewedit{The target selection for Taurus was done as follows. There are 26 stars in \citet[][]{kenyon_pre-main-sequence_1995} that have a visible magnitude $R_{mag}<14$, and a stellar mass between $0.9-3.5 M_\mathrm{Sun}$. We then removed stars that are binaries with a projected separation $<3''$ from this list: seven were identified based on the WDS catalog \citep{Mason2001} and three were first observed with OSIRIS (HBC407, HBC355, and HBC351), but their observations were stopped upon discovery of a bright secondary component in the field of view of OSIRIS. Of the remaining 16 stars satisfying these criteria, we present the observations of the first 12 stars that were targeted. 
For Ophiuchus, there were 28 stars satisfying the same mass and brightness criteria in \citet[][]{Wilking2005}. Seven were found to be binaries based on the literature \citep{barsony_search_2003, cheetham_mapping_2015}. Of the remaining 18 stars, we present the observations of the first 10 stars that were targeted. 
}

Table \ref{table:targets} lists and details the 10 Ophiuchus and \reviewedit{10} Taurus targets, \reviewedit{as well as three non-member targets}, presented in this work. Table \ref{table:science} \reviewedit{describes our} science observations during 2021's first and second Keck observing semesters. We list the seeing measured by Maunkea-DIMM \footnote{\href{http://mkwc.ifa.hawaii.edu/current/seeing/}{http://mkwc.ifa.hawaii.edu/current/seeing/}} in Table \ref{table:skycalib}, as a measure of observing conditions for each night. \reviewedit{As we use a 20-milliarcsecond plate scale, the seeing measurements correspond to $\sim 3 \pm 1$ IFS spaxels, indicating good and stable weather conditions. We experienced unreliable performance of the Keck AO system during some of our observation nights, leading to deteriorated quality of PSFs even in these good conditions.}

\reviewedit{Three stars that are not members of Ophiuchus or Taurus were incorrectly included in our survey. HD 148352 is a foreground interloper \citep{mamajek_distance_2008}, as discussed in more detail in Section \ref{subsec:hd148352}. HBC 353 and HBC 354 could have been erroneously classified as Taurus members in \citet{kenyon_pre-main-sequence_1995}, and \citet{Kraus_2009} list plausible criteria for non-membership due to their under-luminosities and large separations from Taurus's central cloud core. The low parallaxes of HBC 353 and HBC 354, as well as the high parallax of HD 148352, similarly raise concerns about non-membership. We list these targets as ``Non-Member Targets" in Tables \ref{table:targets} and \ref{table:science}. While detection maps and sensitivity curves are presented for these targets, we do not include them in the depth-of-search analysis in Section \ref{subsec:depth_search} and Figure \ref{figure:depth_search}.}

\subsection{OSIRIS data \label{subsec:osiris_data}}
We use the Keck/OSIRIS near-infrared integral field spectrograph ($R \approx 4000$), with the Keck Adaptive Optics system \reviewedit{\citep{Wizinowich2022}} using the bright target itself as the natural guiding star (NGS). We took 3 to 5 sequences of exposures per target in the \reviewedit{Kn5} narrow band ($2.292-2.408\,\mu\mathrm{m}$). The Kn5 filter includes the carbon monoxide bandhead, which is the region of the spectrum with the most spectral features and therefore signal, \reviewedit{due to the presence of \ce{CO} and \ce{H2O} molecules}. Choosing a narrow band filter also doubles the width of the field of view compared to broadband observations. We use the smallest plate scale of 20-milliarcsecond to better spatially resolve the companion from the host star, resulting in a field of view of $0.64''\times1.28''$. Between sequences, we dither by 2-4 pixels in both directions as a way to limit the number of sky observations required. Because a real body would maintain the same relative position to the target star between dithers, this also prevents spatially-dependent instrumental artifacts (such as a bad spaxel) to be misidentified as a planetary signal \reviewedit{(albeit, this does not prevent a background source to be misidentified as a companion)}. 

For each target, we take a sequence of sky background frames with the same exposure time and adaptive optics system off. We limit possible persistence issues for bright targets by acquiring them in alternating sections of the \reviewedit{FOV}. After the first observing run aimed at Ophiuchus, we discovered an issue with detector traces from different lenslets bleeding into one another, rendering part of the field of view on each side of the PSF unusable. This problem is described in more detail in Section \ref{subsec:forward_model}. As a mitigation strategy, we added a 90-degree rotation of the FOV for roughly half of the sequences in subsequent observing runs. 

\begin{figure*}
\gridline{\fig{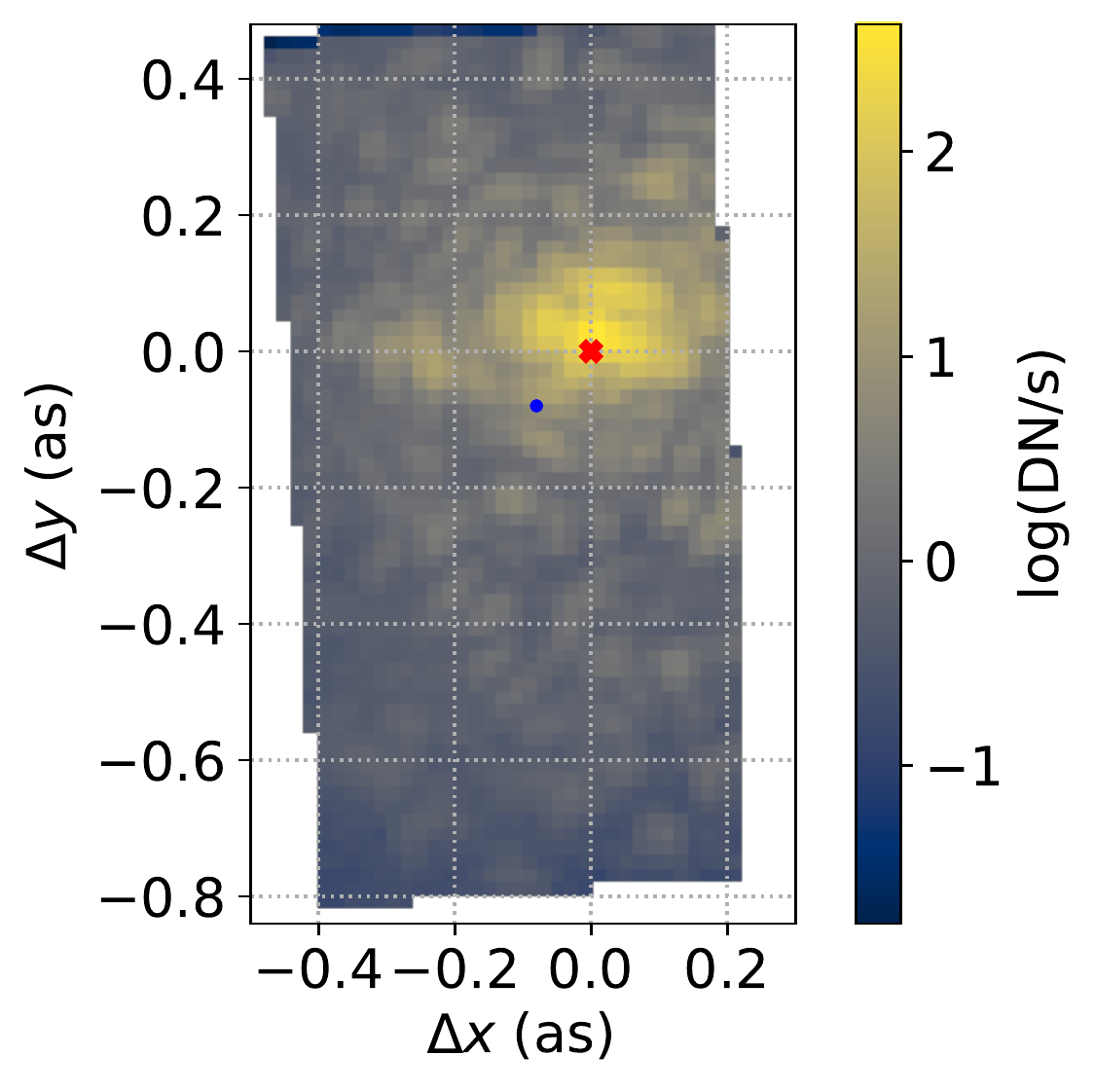}{0.49\textwidth}{(a) Raw image}
		\fig{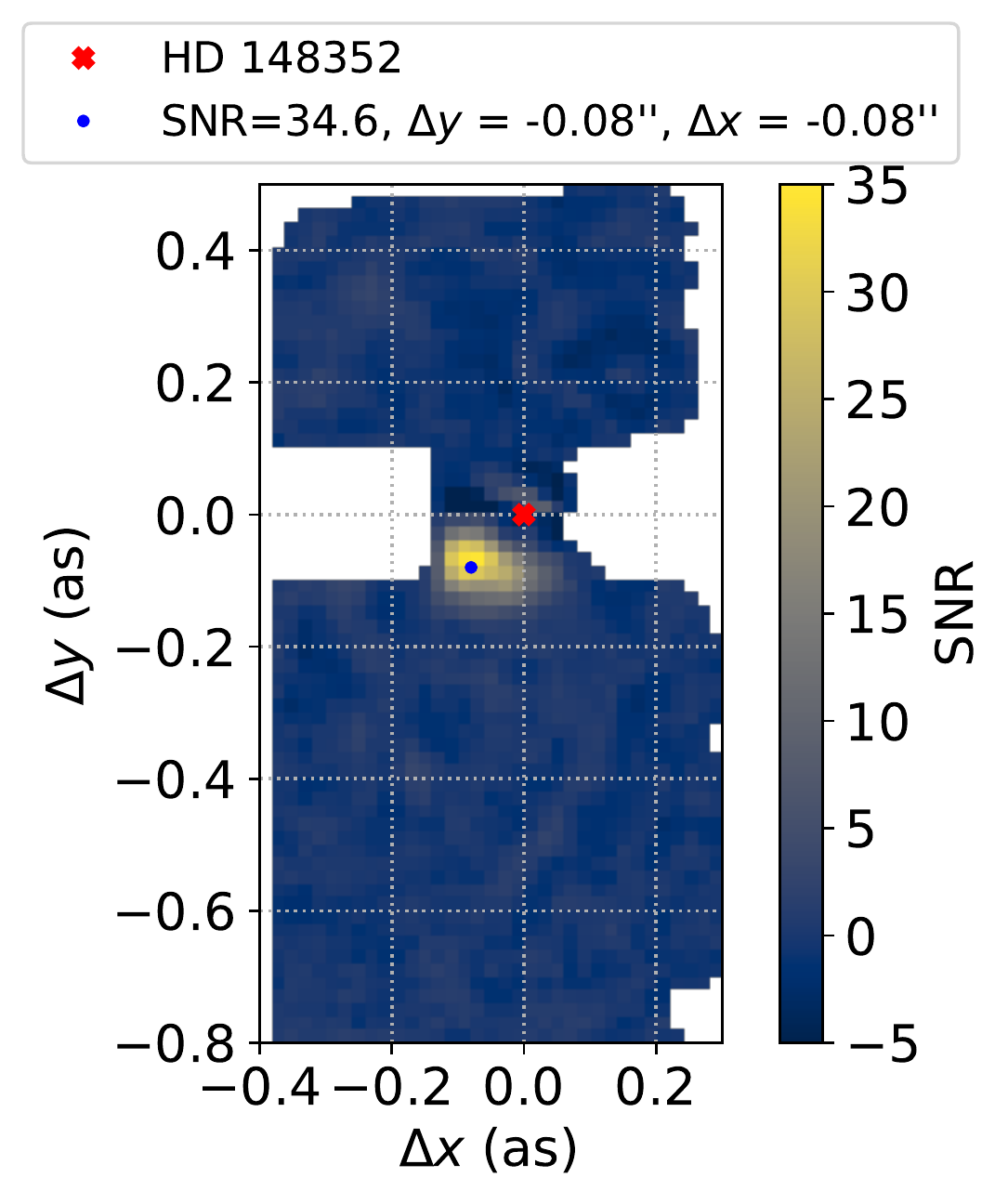}{0.49\textwidth}{(b) S/N map}
          }
  \caption{\label{fig:hd_detection} Detection of a binary companion around HD 148352. (a) The left panel illustrates the point spread function of the host star \reviewedit{(red cross)} from a  median collapsed spectral cube of a 30-second single exposure reduced by the OSIRIS data reduction pipeline. (b) The right panel shows the combined detection map from $\approx 30$ minutes of OSIRIS-Kn5 observations. The signal of the host star has been modeled out and the signal of the companion is highlighted by the blue dot. The S/N $\approx 34$ \reviewedit{detection} at $\Delta y = \Delta x = -0.08$'' from the host star is likely a previously undetected M dwarf binary companion.}
\end{figure*}

The OSIRIS Data Reduction Pipeline \reviewedit{(OSIRIS DRP)} reduces the two-dimensional raw data from the instrument with sky subtraction and bad pixel identification to obtain a three-dimensional datacube, with $N_\lambda = 465$ values along the spectral axis at $n_y \times n_x = 66 \times 51$ spatial locations or spaxels. An example of single exposure collapsed cube is shown in Figure \ref{fig:hd_detection}. \reviewedit{Notably, all of our science data is taken in the field-stabilized mode unlike surveys using ADI, as the high spectral resolution allows us to avoid relying on the angular behavior of speckle noise and planetary signal.}

At the start of each observing night, we take \reviewedit{1 or 2} long exposures (300- or 600-second) images of the sky in the Kn3 narrow band ($2.121-2.229\,\mu\mathrm{m}$), as listed in Table \ref{table:skycalib}. Following the OSIRIS pipeline user manual\footnote{\href{https://www2.keck.hawaii.edu/inst/osiris/OSIRIS\string_Manual\string_v6.pdf}{keck.hawaii.edu/inst/osiris/OSIRIS\_Manual\_v6.pdf}}, we use this to perform a first-order correction of the OSIRIS wavelength solution using \reviewedit{telluric} emission lines due to the \ce{OH-} radical present in the Earth's upper atmosphere, standard values for which are taken from \cite{rousselot_night-sky_2000}. We use the model $\lambda_{\text{standard}} = \lambda_{\text{data}} (1+a_1) + a_0$ to ascertain the constant and linear order offsets $a_0$ and $a_1$. \reviewedit{These constants are determined from the Kn3 narrow-band long exposures and then applied to Kn5 data, as the wavelength correction does not depend on the filter.} The Kn5 skies acquired for the science observations cannot be used for this correction due to the lack of \ce{OH-} lines in this narrow band.

We also collect data from standard A0 stars (which exhibit few spectral features), as listed in Table \ref{table:calibration}, to perform telluric calibration. We extract the stellar spectrum of these targets by aperture photometry as described in \reviewedit{Section \ref{subsec:speckle_model} (without masking a region for a \secondreviewedit{putative} planet)} and divide by a theoretical A0 spectrum model \reviewedit{\citep[PHOENIX synthetic spectra;][]{Husser2013}} to derive the sky transmission for the observations.

\section{Data reduction}
\label{sec:datareduction}

\subsection{\reviewedit{Bad pixels}}
We identify and correct for bad pixels at various steps in our data reduction; all pixels masked are ignored in the final forward modeling fits. After importing the data cubes produced by the OSIRIS DRP, we perform a crude fit to the spectrum at each location in the field of view, using only a forward model consisting only of a simplified starlight component, as described in Section \ref{subsec:forward_model}, \reviewedit{particularly Section \ref{subsec:speckle_model}}; we mark any $>3 \sigma$ outliers in the residuals as bad pixels. We also mask edge pixels in the spectral and both spatial directions, because they often feature artifacts or drops in flux. 
Despite the dedicated additional terms in the forward model as described in Section \reviewedit{\ref{subsec:forward_model}}, we found significantly larger residuals at the core of the deepest telluric lines, which were therefore masked. This is possibly due to the variable line spread function and spectral under-sampling changing the spectrum from spaxel to spaxel.

We also identified a systematic affecting OSIRIS Kn5 filter data for bright targets due to flux between traces on the detector being wrongly assigned, because of the tight arrangement of micro-spectra. Spaxels that are well separated in the sky, but horizontally aligned with the star, are contaminated by the light coming from the on-axis star. We observe the continuum at these extended regions to be much higher at either end of the spectral axis. As the contaminating light corresponds to different wavelengths than the contaminated pixels, modeling this effect is complex, and we instead mask spaxels where the continuum slope is higher than a set factor. This rendered a significant portion of the field of view to be unusable as can be seen in \autoref{figure:SNR_maps}. As a consequence, we improved the observing strategy for Taurus observations \reviewedit{(including HBC 353 and HBC 354)} and rotated our field of view by 90 degrees for roughly half of the data for each target.

\subsection{The framework of \breadsnospace}

\breadsnospace, or the Broad Repository for Exoplanet Analysis, Detection, and Spectroscopy, is a flexible framework that allows forward modeling \reviewedit{of data from} moderate to high-resolution spectrographs. The philosophy of \breads is to have users choose a data class, a forward model function, and a fitting strategy. Data classes normalize the data format, simplifying reduction across different spectrographs while allowing for specific behaviors of each instrument to also be coded into their own specific class.  The forward model (FM) aims to reproduce the data ($d$) as $d=\mathrm{FM}+n$, where $n$ is the noise. The FM is a function of relevant astrophysical parameters of the planet and the host star, but also some \reviewedit{nuisance parameters}. For a general FM within \breadsnospace, \reviewedit{nuisance parameters} do not contain physical information about the planet but are needed to model the data accurately. \reviewedit{\secondreviewedit{For the specific FM used in this work} and discussed in Section \ref{subsec:forward_model}, the linear parameters that are modeling the spurious contribution of the host star, the contribution from telluric-only component, and the contribution from the residual principal components, are all nuisance parameters. On the other hand, planetary characteristics (needed to model its spectrum) such as effective temperature, surface gravity, and radial velocity or its position relative to the star are normal astrophysical parameters, and not nuisance parameters.}

We distinguish between linear and non-linear parameters in \reviewedit{any forward model function used within the \breads framework} because \breads performs an analytical marginalization of all of its linear parameters as described in \reviewedit{\citet{Ruffio2019}} to improve the tractability of the problem\reviewedit{. For} the specific FM used in this work, the contribution from each FM component shown in Figure \ref{figure:forward_model} is a linear parameter.
Indeed, the posteriors for these linear parameters can be calculated analytically without a sampling algorithm like \reviewedit{Markov Chain Monte Carlo (MCMC)}, allowing for increased speed, higher-dimensional or complex models \citep{Ruffio2019, Ruffio2021}.
The definition of a data structure and a forward model leads to the definition of a likelihood assuming Gaussian white noise, which can then be used to either optimize the parameters through a maximum likelihood or derive their posteriors. 

Examples of fitting strategies include a simple grid search optimization, more general optimizers (e.g. Nelder-Mead), or even posterior sampling algorithms like \reviewedit{MCMC}.
The grid search can, for example, be used to compute detection maps or cross-correlation functions by varying respectively the position of the planet or its RV. 

\subsection{\label{subsec:forward_model}Forward model and likelihood}

\begin{figure*}[htpb]
\centering
\includegraphics[width=\textwidth]{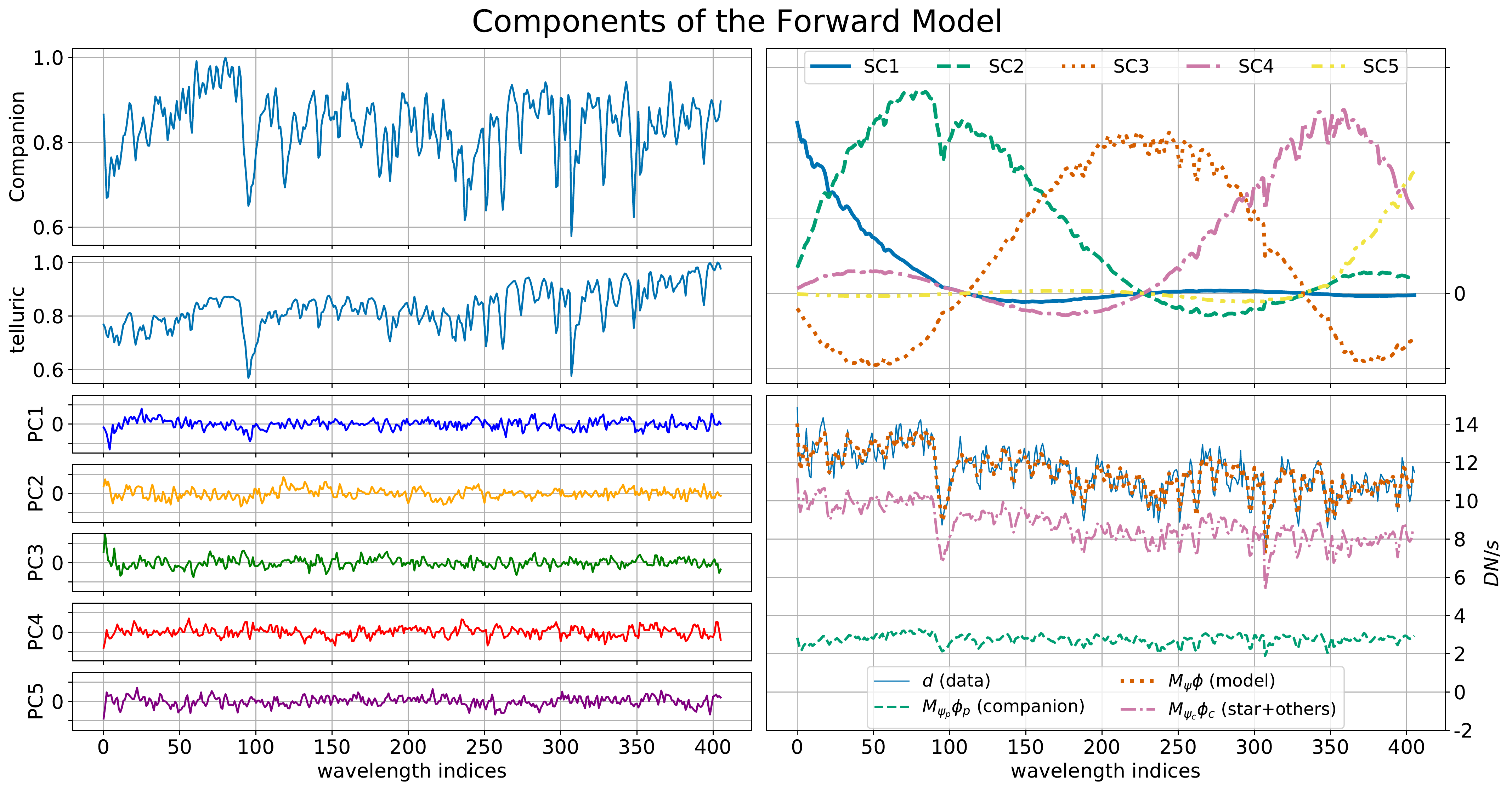}
\caption{\label{figure:forward_model} Schematic of the components of our forward model for a specific spatial location, \reviewedit{used in this work and discussed in Section \ref{subsec:forward_model}}. Other than in the bottom right panel, each spectrum corresponds to a column of the FM matrix $M_\psi$.
Here, we consider a single exposure of the M dwarf HD 148352 B detected in this work (Section \ref{subsec:hd148352}). \reviewedit{Top left: the planet model (Section \ref{subsec:planetmodel}), which is not high-pass filtered and consists of the product of a BT-Settl atmospheric model \citep{allard_model_2010,allard_bt-settl_2014}, and the telluric spectrum (Section \ref{subsec:addmodel}). Top right: the star-light model (Section \ref{subsec:speckle_model}), which is a combination of 10 different spline components to modulate the speckle continuum; we show only 5 in this schematic for clarity; these are labeled SC1, SC2, ..., SC5 (that is, in the model we use, we have 10 components SC1, SC2, ..., SC10). Bottom left: the principal components of the residuals (Section \ref{subsec:addmodel}) that are obtained from a preliminary fit over a disjoint area. Bottom right: a linear combination fits the spectral data at the spaxel.}}
\end{figure*}

We use the formalism developed in \citet{Ruffio2019, Ruffio2021} to build a forward model of the data. In \citet{Ruffio2021}, the forward model included three terms: the planet signal, the starlight, and principal components of the residuals, all of which were high-pass filtered. The main difference in this work is that the continuum of the spectra is included in the forward model and the speckle modulations are modeled with a spline.

The model is defined as $d= M_\psi \phi + n$, with $d$ being a vector with $b\times b\times N_\lambda)$ elements and $n$ being a Gaussian \reviewedit{random} noise vector \reviewedit{with zero mean} of the same size. \reviewedit{The model matrix $M_\psi$ is determined by non-linear parameters $\psi$, independent of the linear parameters $\phi$.} We consider data within a $b\times b$ box around a spatial location, with a side length of $b=3$ pixels. Before fitting, $d$ is scaled by the \reviewedit{standard deviation of the random vector}, following $d \rightarrow d / s$ element-wise. This \reviewedit{standard deviation vector} $s$ is set to the highest of two terms at each pixel: first, the square root of the continuum obtained by taking a moving average in spectral direction, which is roughly proportional to the photon noise, and second, a chromatic background noise floor. The noise floor is calculated in each slice by only considering half of the pixels with the lowest continuum value in the field of view, as a way to not include the central star, and taking their standard deviation after subtracting the continuum in each spectrum. The model $M_\psi \phi$ is similarly scaled by the noise $s$. Noise scaling gives a statistically accurate likelihood assuming uncorrelated Gaussian noise. We analytically marginalize over the linear parameters $\phi$ (planet or starlight continuum) after computing and optimizing $M_\psi$ over the non-linear parameters $\psi$ (astrometry, radial velocity, effective temperature).

\reviewedit{Within the framework of the forward model used in this work, the noise vector $n$ is assumed to be uncorrelated and Gaussian, which results in a Gaussian distribution of residual signal-to-noise ratio (S/N). Section \ref{subsec:snr_maps} and Figure \ref{figure:hist_snr} discuss the non-Gaussian behavior of the noise, which induces non-Gaussianity in the distribution of S/N residuals. We correct this by setting a detection threshold that gives us a false positive rate equivalent to a $5\sigma$ threshold for a Gaussian S/N  distribution.}

The model $M_\psi \phi$ is divided into four components, which are described in more detail in the following sections \reviewedit{and shown in Figure \ref{figure:forward_model}: a planetary signal $M_p$ (1 parameter $\phi_p$; Section \ref{subsec:planetmodel}), starlight $M_\mathrm{star}$ ($ 3\times 3 \times 10$ parameters $\phi_\mathrm{star}$; Section \ref{subsec:speckle_model}), a telluric-only model $M_\mathrm{tel}$ (1 parameter $\phi_\mathrm{tel}$; Section \ref{subsec:addmodel}), and residual principal components $M_\mathrm{res}$ (5 parameters $\phi_\mathrm{res}$; Section \ref{subsec:addmodel})} such that:
\begin{equation}
     M_\psi \phi = M_p \phi_p +  M_\mathrm{star} \phi_\mathrm{star}+  M_\mathrm{tel} \phi_\mathrm{tel}+  M_\mathrm{res} \phi_\mathrm{res}.
\end{equation}
Each term composes different columns in $M_\psi$, with the different rows corresponding to spaxels or wavelengths. The $\phi$ vector \reviewedit{therefore contains a total of $97$ linear parameters} that define the contributions from these different components.

\subsubsection{\label{subsec:planetmodel} The planet model}
The first component, \reviewedit{shown in the top left panel of Figure \ref{figure:forward_model},} models the planetary signal. For each exposure, we independently extract the associated PSF by taking the $b \times b$ spatial region centered at the star, normalizing the chromatic slices of data, and scaling them by a theoretical planetary signal spectrum multiplied by a telluric model. \reviewedit{The telluric model is extracted from data of standard A0 stars (Section \ref{subsec:osiris_data}) and is the same as the one used in Section \ref{subsec:addmodel}.} \reviewedit{For the theoretical spectrum, we} use a \texttt{BT-Settl-CIFIST2011c} atmospheric model \reviewedit{\citep{allard_model_2010,allard_bt-settl_2014}}, which include one-dimensional cloud models that consider timescales of condensation, coalescence, gravitational settling, and mixing, as well as mixing length theory and non-equilibrium chemistry \reviewedit{\citep{freytag_role_2010}}. We start with a spectrum with characteristics for a nominal directly imaged companion, broaden absorption lines \reviewedit{to} Keck/OSIRIS's resolution, and shift wavelengths by $1 - (RV-RV_{\textrm{bary}}) / c$, with $RV_{\textrm{bary}}$ the barycentric RV. We scale the planet model to the total measured flux of the star in each exposure. Thus, the retrieved parameter for the flux of the planet is in units of stellar flux.

If $\phi_p$ is the coefficient that multiplies the planet component in $M_\psi \phi$, with $\phi=[\phi_p,\dots]$, then $\phi_p$ quantifies the estimated amount of planetary signal as a planet-to-star flux ratio. 
We use the formalism developed in \reviewedit{\citet{Ruffio2019} and \citet{Ruffio2021}}, where if $\sigma_p$ is the estimated uncertainty of $\phi_p$, then $\phi_p / \sigma_p$ gives the signal-to-noise ratio for a fixed planet effective temperature, surface gravity, position, and radial velocity. 

\subsubsection{\label{subsec:speckle_model} The speckle model}
The second component, \reviewedit{shown in the top right panel of Figure \ref{figure:forward_model},} models the starlight, also known as speckles, independently for each spaxel. In existing studies, the speckle noise is often reduced by high-pass filtering the data, or by normalizing the continuum. High-pass filtering can have undesirable effects due to non-linearities of the filter and edge effects, for example when using a typical median filter with a sliding window. Not subtracting the continuum can also help retain more information about the atmosphere of the planet. 
This is why we propose to fit the continuum of the speckles with a spline jointly with the planetary signal.
Spline interpolation is appropriate for modeling the chromatic continuum of data because the point spread function scales with wavelengths. Specific speckles move outward with wavelength and therefore modulate the continuum of the starlight at a specific location as they move across it. This is well modeled by a spline with low-degree polynomials.

\secondreviewedit{Ten nodes are placed along the spectral direction.} The density of nodes is set higher where the continuum is steeper. \reviewedit{Regions of the spectrum with steep variations have large derivatives; so, the cumulative sum of the absolute values of the derivatives increases faster in regions with a steeper continuum. we start with the nodes exactly equispaced in the spectral direction, compute the positions that equally divide the cumulative sum of the absolute value of the gradient of the spectrum at a given spaxel, and drift our nodes 30\% towards these positions.} We construct a spline interpolation model, with 10 piece-wise cubic polynomials modeling the continuum between each pair of nodes. For our narrow band spectra, ten is an appropriate choice for the number of pieces, as it is high enough to model the modulation due to the chromatic scaling of the PSF but not too high such that the continuum model starts fitting planetary absorption features. We multiply these model sub-components by the stellar spectrum to get the starlight model. This star spectrum is extracted from the same science exposure \reviewedit{using aperture photometry,} by summing flux within a rectangular aperture around the host star in each chromatic slice, after masking a $b\times b$ spaxel region around the \secondreviewedit{putative} planet to prevent self-subtraction of any planetary signal and improve sensitivity. In summary, the starlight model is a linear combination of the model contributions for each node, which is illustrated in Figure \ref{figure:forward_model} (components in the top right linearly combine to give the starlight model in the bottom right).

\subsubsection{Additional model components \label{subsec:addmodel} }

The third component, \reviewedit{shown second from top on the left half of Figure \ref{figure:forward_model},} is similar to the planet model but assumes a flat planet spectrum and therefore only includes the telluric lines, \reviewedit{extracted from aperture photometry of standard A0 stars as discussed in Sections \ref{subsec:osiris_data} and \ref{subsec:speckle_model}}. This component is necessary to avoid the planet model from being used to fit the starlight contribution. Indeed, the signal-to-noise ratio of the starlight model is generally much lower than the telluric standard used in the planet model. 
This occasionally leads the fit to incorrectly use the planet model, which includes telluric lines, to preferentially fit the speckles instead of using the starlight model. This is because the added noise from the starlight spectrum can be worse than the extra residuals from the exoplanet atmospheric lines when no planet is present. Including a component with telluric lines in the model mitigates this problem.

The final fourth component, \reviewedit{shown in the bottom left panel of Figure \ref{figure:forward_model},} consists of principal components of the residuals of a preliminary fit to account for any other effect that could have been neglected in modeling the starlight. For example, the resolution of the instrument is known to be field-dependent so it could explain some variability of the spectra between spaxels.
We perform a fit for two small portions \reviewedit{(10 spaxels $\times$ 10 spaxels)} of the FOV spatially \reviewedit{directly} above and below the star and compute the eigenvectors of the residuals' covariance. We add this orthonormal principal component basis as the fourth forward model component, for the modeling of the bottom and top (respectively) halves of the full FOV. Using a disjoint portion of the sky to calculate the principal components ensures that the planetary signal is not subtracted away in the subsequent fit.

\subsection{\label{subsec:planet_search} Planet search}

In this section, we describe the calculation of combined S/N maps for each target, which are used for planet detection, and the derivation of the \reviewedit{companion} sensitivity using simulated planet injection and recovery.
 
We set the values of the non-linear parameters to $RV = 0\,\mathrm{km/s}$, effective temperature $T_\text{eff} = 1700$ K, and $\log g = 3.5$. This definition of the model is discussed in Section \ref{subsec:model_choice}. Iterating over different spatial locations in the field of view, we construct the forward model $M_\psi$ as described in Section \ref{subsec:forward_model}. For each exposure, we fit for the linear parameters $\phi$ over this grid of spatial shifts relative to the host star and barycentric corrected RV values. The values of interest are the estimated flux ratio of the planet $\phi_p$ and its associated uncertainty $\sigma_p$. The S/N is defined as $\phi_p/\sigma_p$ and the sensitivity of the observation is related to $\sigma_p$. As detailed below, both the S/N and the sensitivity derived from the FM fit need to be calibrated to account for systematics, and then all the frames combined to produce a single two-dimensional S/N map. We expect a real astrophysical source to have an RV of under a few tens of km/s, which is smaller than the spectral resolution of OSIRIS, so the S/N should peak for models with $RV\approx0\,\mathrm{km/s}$.

\reviewedit{The desired theoretical S/N, derived from the maximum likelihood statistical framework, assumes white Gaussian noise and no algorithmic systematics; however, the S/N calculated by the above methodology is generally overestimated. So, we empirically calibrate the S/N. A correctly calibrated radial velocity cross-correlation function (CCF) is expected to have a zero median and unit standard deviation.} We first compute it for 41 RV values from -4000 to 4000 km/s. We subtract the median of the S/N values and normalize them by the standard deviation calculated over the RV direction to ensure a unit standard deviation of the S/N. We then remove outliers in the temporal direction using a 5-sigma clipping based on the median absolute deviation. The different exposures are combined using a weighted mean such that
\begin{align}
\phi_p &= \Sigma_j \frac{\phi_{p, j}}{\sigma_{p, j}^2} \Bigg{/} \Sigma_j \sigma_{p, j}^{-2} \label{equation:addflux} \\
\sigma_{\phi_{p}} &= \Big{(} \Sigma_j \sigma_{p, j}^{-2} \Big{)}^{-1/2} 
\label{equation:combine_frames}
\end{align}
where $\sum_j$ represents summing \reviewedit{over exposures} for a given target. Finally, the S/N map for a null barycentric corrected RV is once again normalized by its standard deviation. For the Taurus targets, a subset of the individual exposures was first rotated by ninety degrees before combining to account for the field of view rotation. These final S/N maps (\reviewedit{Figure} \ref{figure:SNR_maps} in \reviewedit{Appendix \ref{app:detecmaps}}) are the ones used for planet detection.

Additionally, as is routine in high-contrast imaging, we calibrate for the algorithmic throughput using injection and recovery tests. For every single frame, we inject a fake exoplanetary signal on integer pixels at varying angular separations from the target star and recover it using our forward model approach. We interpolate the algorithmic throughput between the discrete values of separation in pixels using a piecewise polynomial spline.

An example of companion detection is shown in Section \ref{subsec:hd148352}. The definition of the detection threshold is discussed in Section \ref{subsec:snr_maps}. It is a function of the empirical distribution of S/N residuals and is set to $8\times\sigma_p$ in this work.

\subsection{\label{subsec:model_choice} Effect of the choice of the model spectrum}

The effective temperature $T_\text{eff}$, radial velocity $RV$, or the specific gravity $\log g$ determine the companion spectral model and are fixed when generating these detection maps or sensitivity curves. In this section, we study the effect of a $T_\text{eff}$ mismatch between the model spectrum and a real planet using simulated planet injection and recovery similar to the calculation of the algorithmic throughput described previously.
The main goal of this section is to select an appropriate standard value for $T_\text{eff}$ used to generate detection maps, so that our method with this specific $T_\text{eff}$ is sensitive to planets with a wide range of real $T_\text{eff}$'s. 

\begin{figure}[htpb]
\centering
\includegraphics[width=\columnwidth]{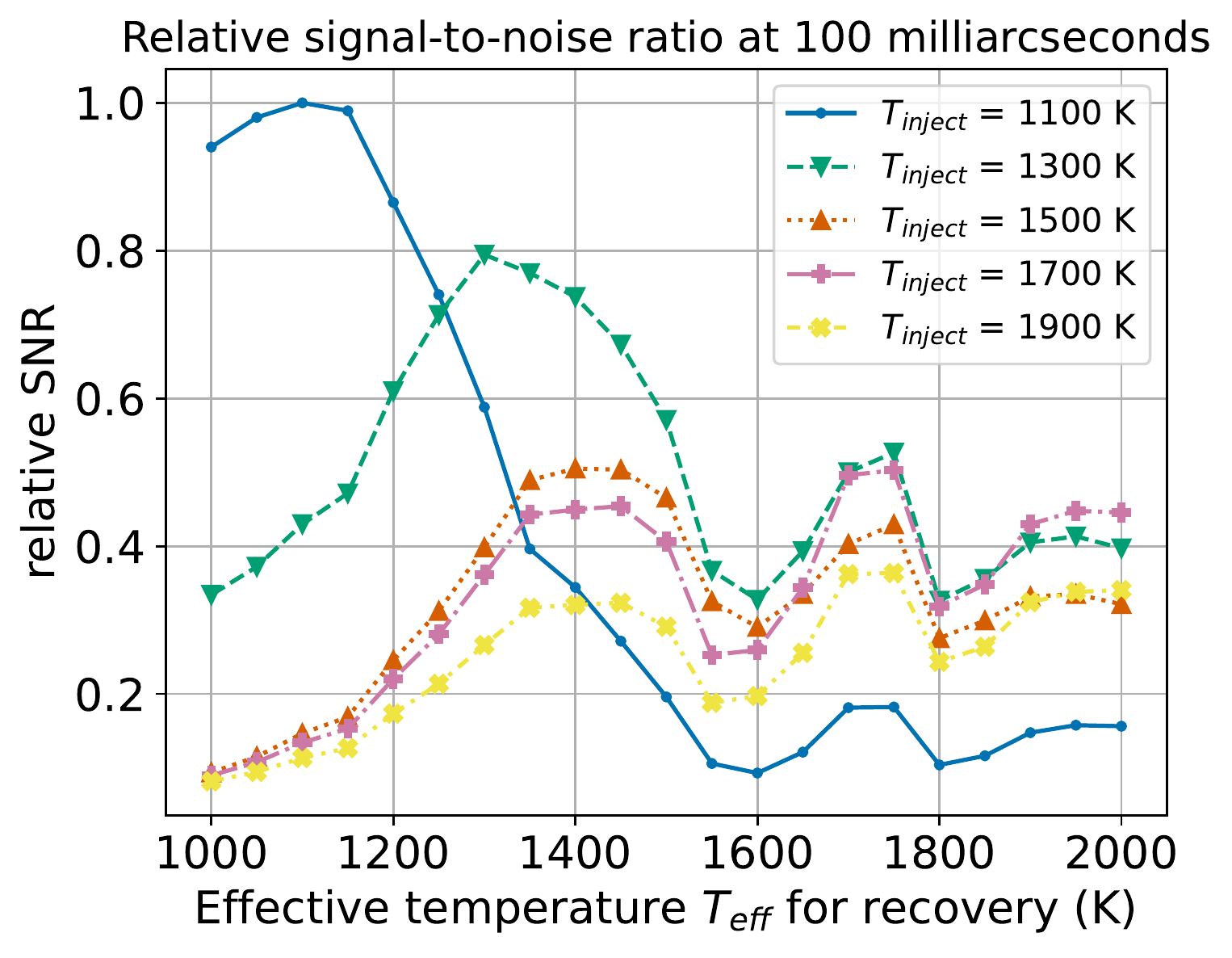}
\caption{\label{figure:temp_recover} Relative signal-to-noise ratio as a function of the true effective temperature of the simulated planets (colored curves) and the model temperature ($x$ axis) used to recover it. Simulated planets are injected at a separation of 100 mas. } 
\end{figure}

At a fixed angular separation of 0.1'' from the star, we independently inject several simulated planetary signals in the data with a particular $T_\text{eff}$ and attempt to recover it using a model built from a range ($1,000-2,000\,\mathrm{K}$) of other effective temperatures. We can therefore plot the S/N of a planet as a function of its true effective temperature and the one used in the model as shown in Figure \ref{figure:temp_recover}. 

A temperature of $1700\,\mathrm{K}$ appears adequate to recover a wide range of temperatures, especially at the higher range. A lower temperature would not help as detectable planets and brown dwarfs in Ophiuchus and Taurus would not have had time to cool to cooler atmospheres within a few million years. However, this plot is not specific to the targets presented in this work and can therefore be used for older cooler planets too. For example, if targetting older stars, the second set of reductions using a lower effective temperature would be necessary to detect planets with $T_\text{eff} < 1400\,\mathrm{K}$. The degeneracy between planet signals with $T_\text{eff}$ = 1500, 1700, and 1900 K, resulting in the multi-modal curves is due to the limited bandwidth of the filter as the spectra at these different temperatures are similar in the narrow spectral range.

\reviewedit{For a general case, a similar procedure as above can be used to study mismatches between other parameters that set the planetary spectrum used in the FM; for instance, surface gravity $\log g$ for the FM described in this work. For our dataset, however, the choice of $\log g$ predominantly determines the spectral energy distribution (SED) and does not affect the molecular \ce{CO} lines in our narrow-band Kn5 spectra. Inversely, a broader spectral range (obtained using the broadband K-band filter, for example) is better for constraining $\log g$. As our work focuses on detection and not characterization, the narrow-band OSIRIS Kn5 filter is a good choice because it offers a larger FOV. In hindsight, detector traces bleeding between different lenslets (in the Keck/OSIRIS Kn5 observing mode) makes the Kn5 filter a poorer option, without a 90-degree rotation strategy.} 

\reviewedit{We additionally note that, for detection, it is not critical for the model to perfectly mimic the expected signal, as long as it includes the correct molecular lines. \ce{CO} and \ce{H2O} lines are present in all of our spectral models across this temperature range, and so the strength of detection is driven by the depth of molecular lines in the signal itself.} 

\subsection{Code demonstration: detection of a binary \label{subsec:hd148352}}

We describe the detection of a stellar companion with a flux ratio of $\approx 0.92\%$ around HD 148352 at a separation of $\approx 110$ mas. We use this as a validation of the technique at close separations, as well as a demonstration of the further characterization possible with our approach and \breadsnospace. HD 148352 was listed in \citet{Erickson2011} as a possible young star in the region of Ophiuchus due to X-ray emission, which led us to incorrectly include it in our survey. Proper motion of HD 148352 (values for projected RA and declination $\mu_{\alpha\cos\delta} \approx -52$ mas/yr, $\mu_{\delta}\approx -64$ mas/yr) is larger than Ophiuchus stars (at $\approx 140$ pc, with $\mu_{\alpha\cos\delta} \approx -10$ mas/yr, $\mu_{\delta}\approx -27$ mas/yr), with values consistent with a foreground F dwarf \citep{mamajek_distance_2008}. As the target is unrelated to the star-forming cluster and given \reviewedit{F stars have a binary fraction $>50\%$} \citep{Fontanive2018}, it is likely a stellar binary companion.

\begin{figure}
\centering\includegraphics[width=\linewidth]{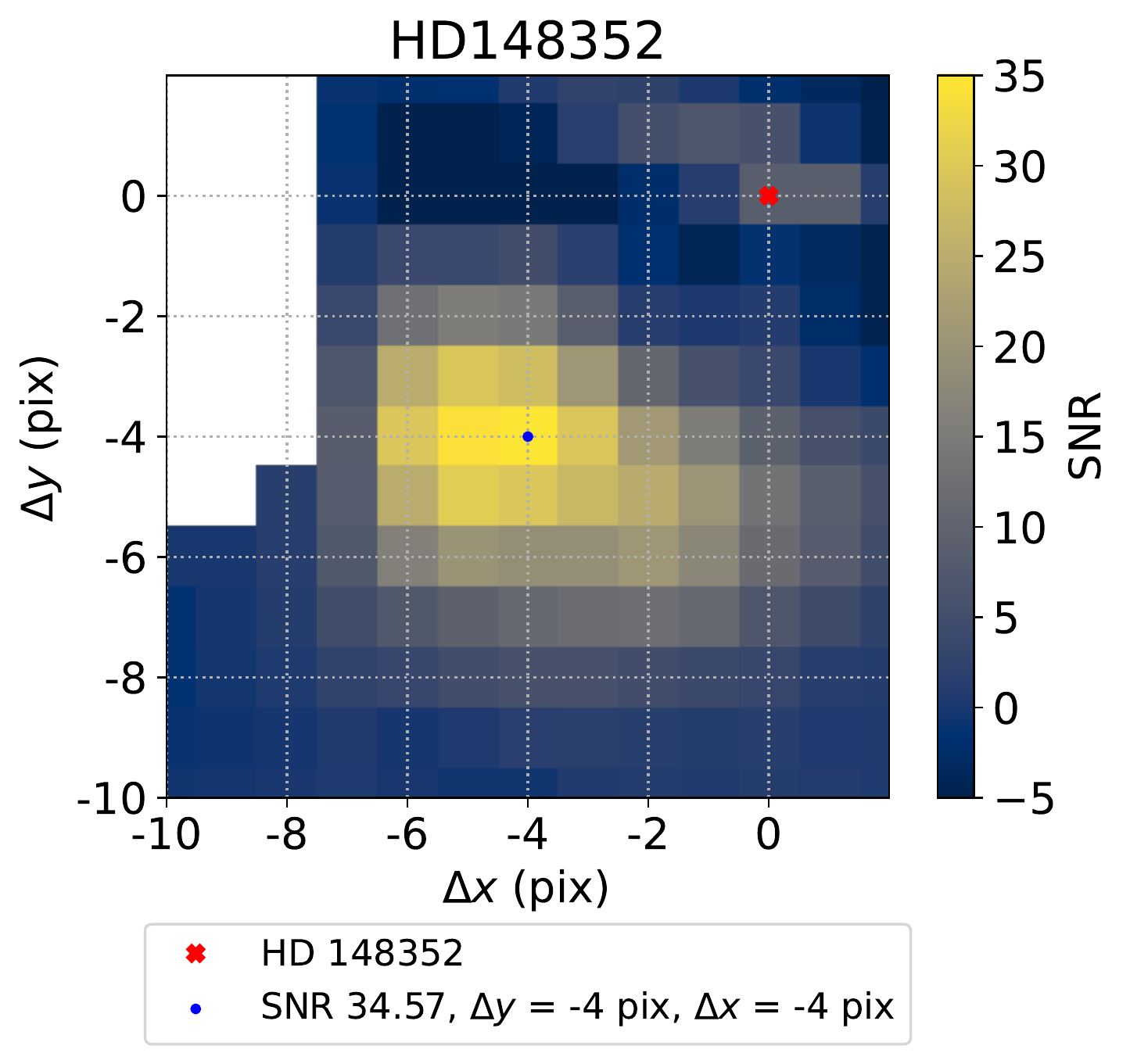}
  \caption{\label{fig:hd_detection_zoom} Zoomed-in section, of the combined detection map from Figure \ref{fig:hd_detection}, constructed from $\approx 30$ minutes of OSIRIS-Kn5 observation around HD 148352. The position with the highest S/N is $\Delta x = \Delta y = 4$ pix away from the star, with a 0.02" plate scale.}
\end{figure}

\begin{figure}[bth]
\centering
\includegraphics[width=\columnwidth]{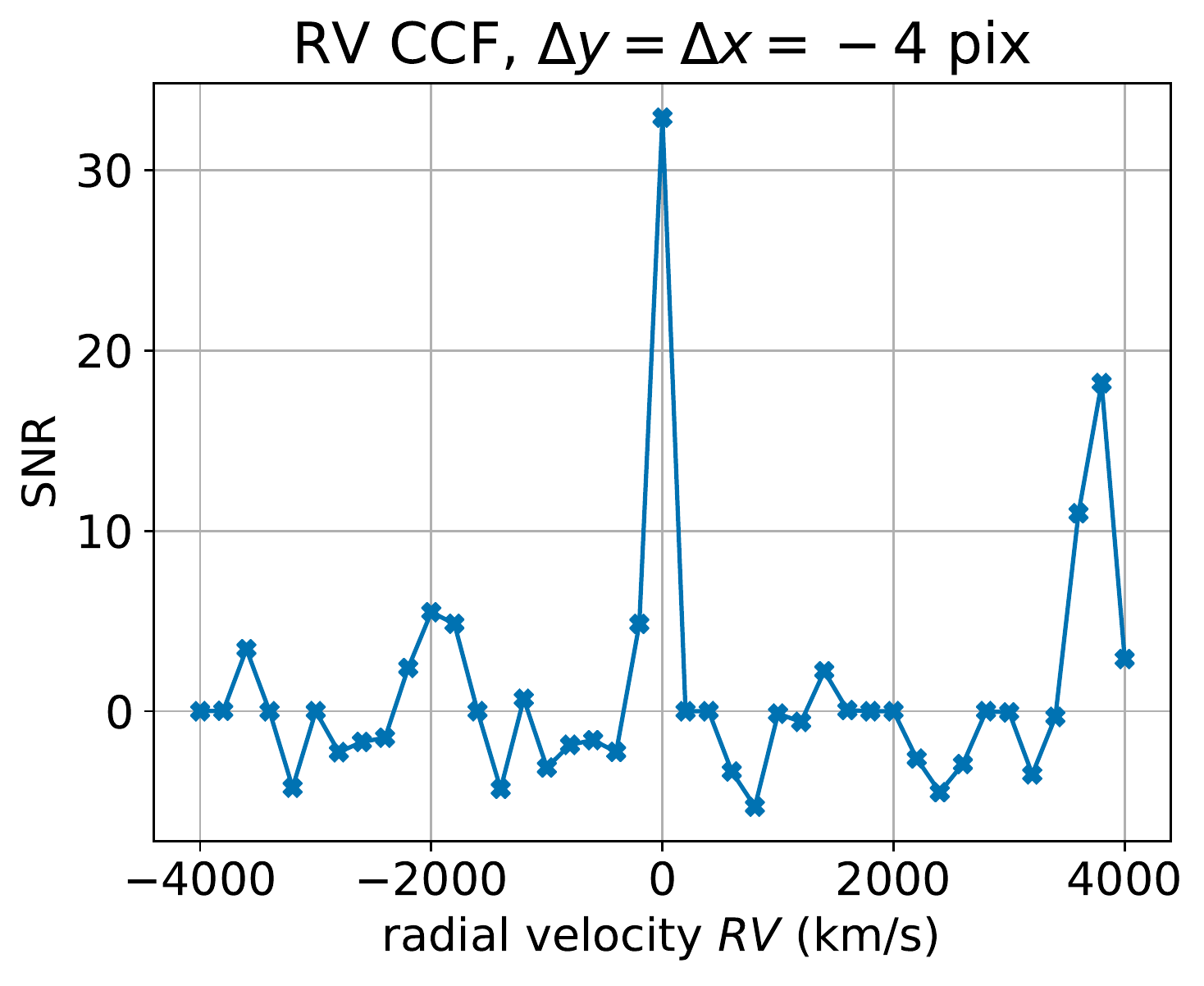}
\caption{\label{figure:hd_rvccf} Radial Velocity Cross-Correlation Function for the detected binary, up to a constant scaling factor. We see a peak signal-to-noise around \reviewedit{RV $\sim$ 0 km/s}, which is expected for a real astrophysical signal. The supplementary peak at $\sim3700$ km/s corresponds to the wavelength shift between two CO band-heads at 2.294 $\mu$m and 2.323 $\mu$m (visible in the bottom right panel of Figure \ref{figure:forward_model}).}
\end{figure}

Figures \ref{fig:hd_detection} and \ref{fig:hd_detection_zoom} show the detection map constructed as in Section \ref{subsec:planet_search} from roughly 30 minutes of data. For the initial planet search, we look for an exoplanetary signal with $RV =0\,\mathrm{km/s}$, $T_\text{eff} = 1700$ K, and $\log g$ = 3.5. We note an elliptical feature of peak S/N $\approx 34$ at the spatial shift of $\Delta y = \Delta x = -4$ spaxels.

Figure \ref{figure:hd_rvccf} shows an RV cross-correlation function \reviewedit{(CCF)} at this location, where we plot S/N \reviewedit{defined in Section \ref{subsec:planet_search}} as a function of the RV used in the forward model, with a peak around zero RV. \reviewedit{Notably, to compute this RV CCF, we do not use a high-pass filter or normalization to remove the continuum of the spectrum. Instead we still forward model the diffracted starlight and planetary signal together, with a spectral shift in the latter corresponding to each radial velocity value.}

\begin{figure}[bth]
\centering
\includegraphics[width=\columnwidth]{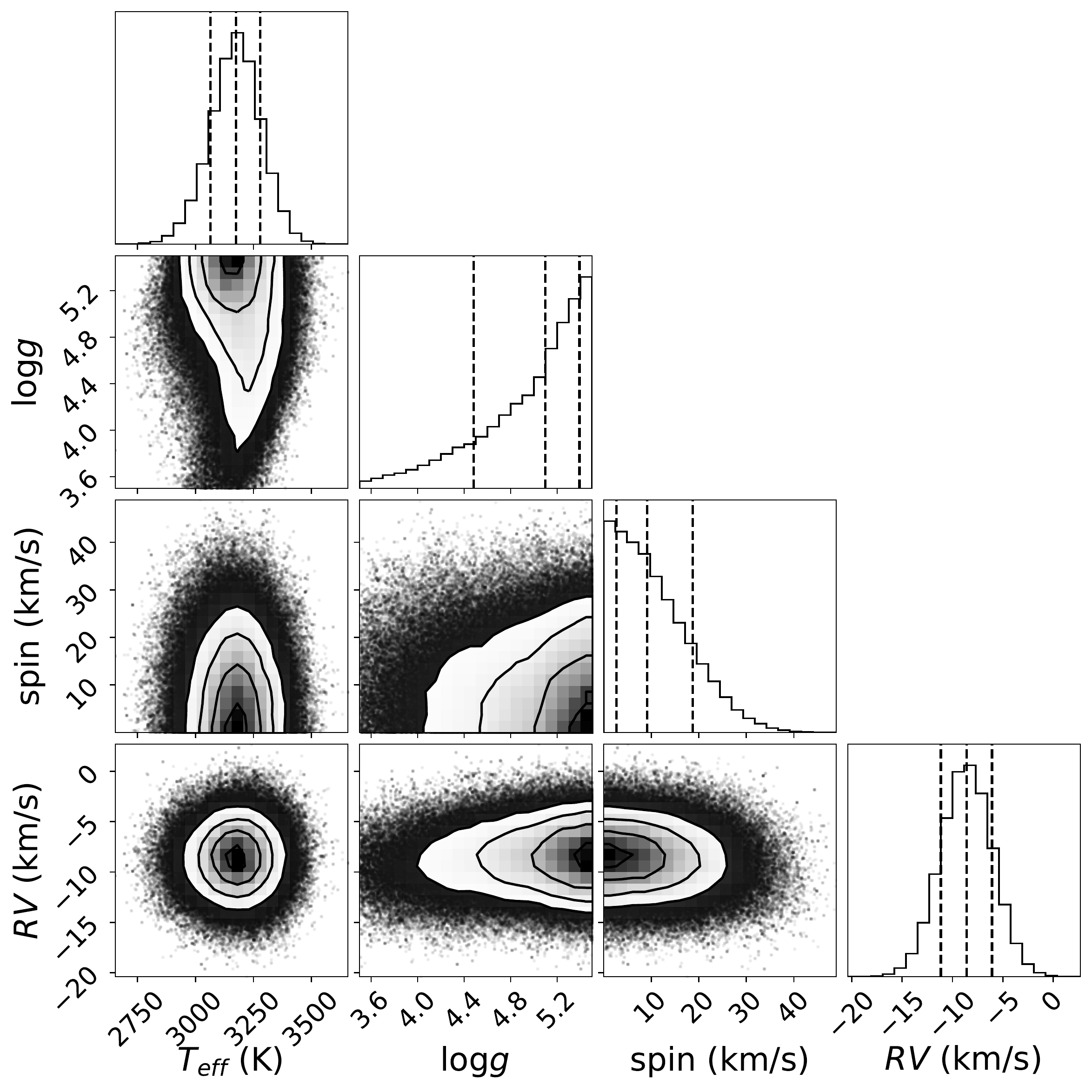}
\caption{\label{figure:hd_corner} Corner plot showing posteriors for $T_\text{eff}$, $\log g$, spin, and RV for detected M dwarf companion to HD 148352, from a single 30-second exposure. The best-fit parameters are $T_\text{eff}=3200\pm100\,\mathrm{K}$, $\log(g)>4$, $\mathrm{spin} <25\,\mathrm{km/s}$, and $RV =-8.6\pm2.5\,\mathrm{km/s}$. The lower and upper limits are given for a 95\% probability. We use an \texttt{emcee} samplers with 512 walkers, 1000 burn-in, and 1000 real samples. \breads can similarly characterize possible exoplanetary detections.}
\end{figure}

To generate detection maps, we set the values of all other non-linear parameters $\psi$ ($RV, T_\text{eff}, \log g$), and optimized over the spatial coordinates $y$ and $x$. To further characterize this signal, we fix the spatial shift $\Delta y = \Delta x = -4$ pix for our forward model, and vary effective temperature ($T_\text{eff}$), spin, radial velocity $RV$, and surface gravity ($\log g$) in $\psi$. The \texttt{BT-Settl-CIFIST2011} includes spectra models for varying $T_\text{eff}$ and $\log g$, and we broaden or shift the spectral lines appropriately in the wavelength space for given $RV$ and spin of the companion. For this higher-dimensional optimization over $\psi$, we use a \texttt{emcee} Markov Chain Monte Carlo sampler instead of a grid search. 

Figure \ref{figure:hd_corner} shows posteriors for these spectral parameters derived from a single frame of data. Based on its $T_\text{eff} = 3200 \pm 100$ K, we believe that the binary companion is an M dwarf \citep{morrell_exploring_2019}. \reviewedit{Table 5\footnote{\href{https://www.pas.rochester.edu/~emamajek/EEM\string_dwarf\string_UBVIJHK\string_colors\string_Teff.txt}{pas.rochester.edu/$\sim$emamajek/\\EEM\_dwarf\_UBVIJHK\_colors\_Teff.txt}} of \citet{2013Pecaut} lists $T_\text{eff}$ of M dwarfs; our posterior $T_\text{eff}$ is closest to the prediction for an M4V dwarf. We perform synthetic photometry with M and L dwarf spectra, from the SpeX Prism Library \citep{2017Burgasser}, to compute the (K - Kn5) color for the 2MASS K-band filter; the color term's dependence on spectral type is modeled as a second-order polynomial to get a correction of (K - Kn5) $\sim$ 0.1 for M4V dwarfs. With HD 148352's distance and absolute $K_\text{mag}$ as $1/p \sim 76.43$ pc and 6.511 respectively and a flux ratio of the companion at $0.92\%$, we get a final corrected apparent and absolute $K_\text{mag}$ for HD 148352 B as 11.7 and 7.3. This is comparable to the $M_\text{K$_s$}$ \citet{2013Pecaut} list (7.36) for an M4V dwarf.} Our data from HD 148352 is from a narrow K band, and more observations with a wider spectral coverage (say, using a broadband filter) will make this estimate more reliable.

\begin{figure}[bth]
\centering
\includegraphics[width=\columnwidth]{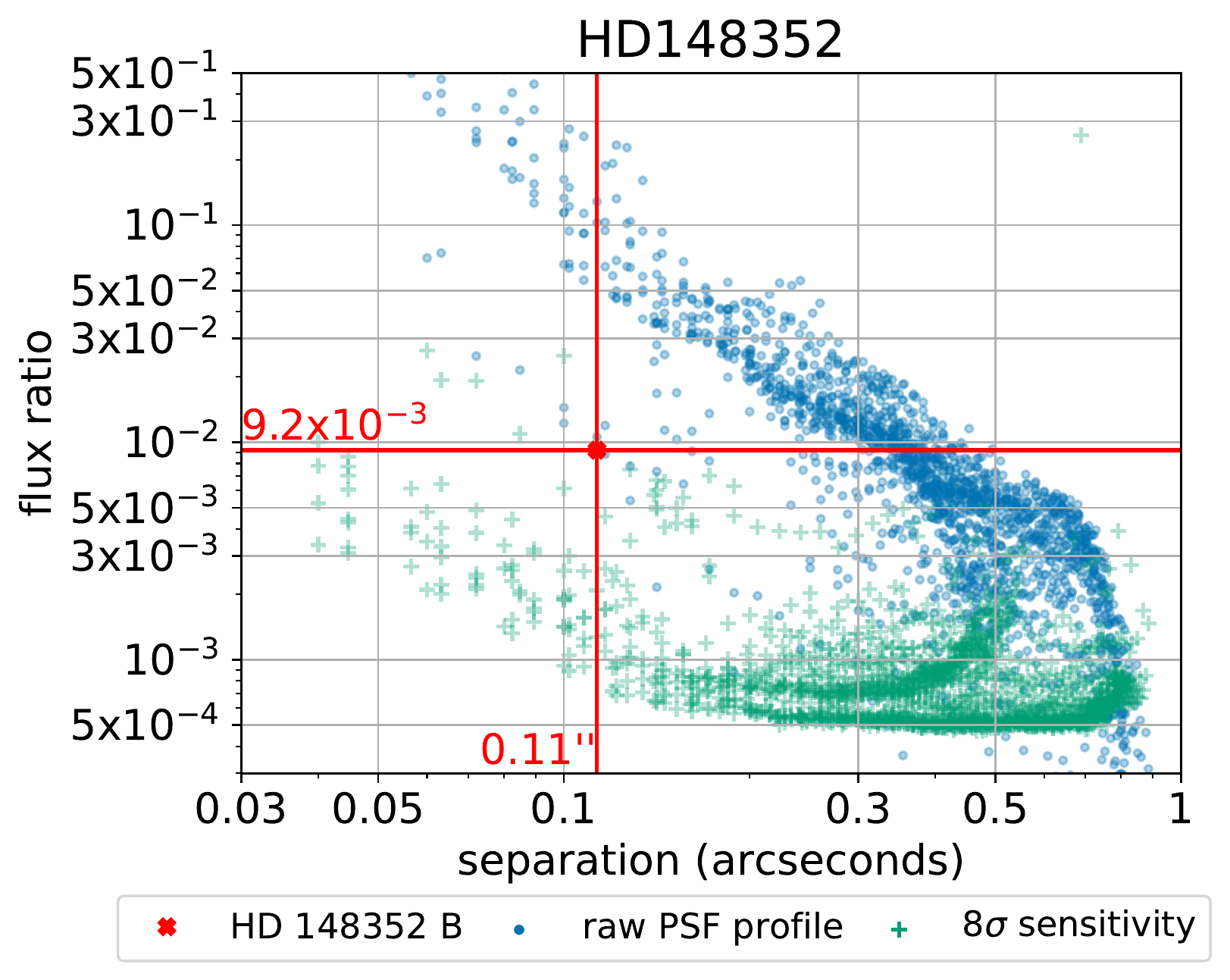}
\caption{\label{figure:hd_contrast} Sensitivity curve (green) at $8\times\sigma_p$ for HD 148352 as a function of angular separation from the host star, overlaid with the raw point spread function profile of the host star (blue) which quantifies speckle intensity. Each data point corresponds to an individual spaxel in the combined detection map. As discussed in \ref{subsec:snr_maps}, the choice of this detection threshold \reviewedit{($8\sigma_p$)} accounts for the non-Gaussianity of the S/N residuals and is equivalent to $5\sigma$ assuming Gaussian noise.
Based on the best-fit parameters from Figure \ref{figure:hd_corner}, the model parameters were specifically set for this \reviewedit{plot} to $RV = -8.6\,\mathrm{km/s}$, $T_\text{eff} = 3200$ K, and $\log g = 5$. The detected M dwarf (red) is about an order of magnitude brighter than our sensitivity limit and 1-1.5 orders of magnitude lower than the speckle noise. The sensitivity curves raise up at wider separations as we effectively have fewer images at the edge, due to dithering between sequences of images.}
\end{figure}

Using these best-fit parameters of $RV \approx -8.6\,\mathrm{km/s}$, $T_\text{eff} \approx 3200$ K, and $\log g \approx 5$, we get $35\sigma$ characterization S/N for a companion with a flux ratio relative to the host star of $\sim0.92\%$. 
In Figure \ref{figure:hd_contrast}, we plot this flux ratio and sky separation of the M dwarf, along with a curve mapping the $5\sigma$ sensitivity (S/N = 8 $\leftrightarrow$ $8\sigma_p$ level, based on Section \ref{subsec:snr_maps}) of our forward model approach at these best-fit parameters, as well as the PSF profile for HD 148352. 
We detect a $<1\%$ companion well below the speckle noise due to the diffracted starlight and are sensitive to planets $\sim 100$ times fainter than the speckles at $\sim 100$ mas. 

\section{Results}
\label{sec:results}

\subsection{\label{subsec:snr_maps} S/N maps}

\begin{figure*}
\gridline{\fig{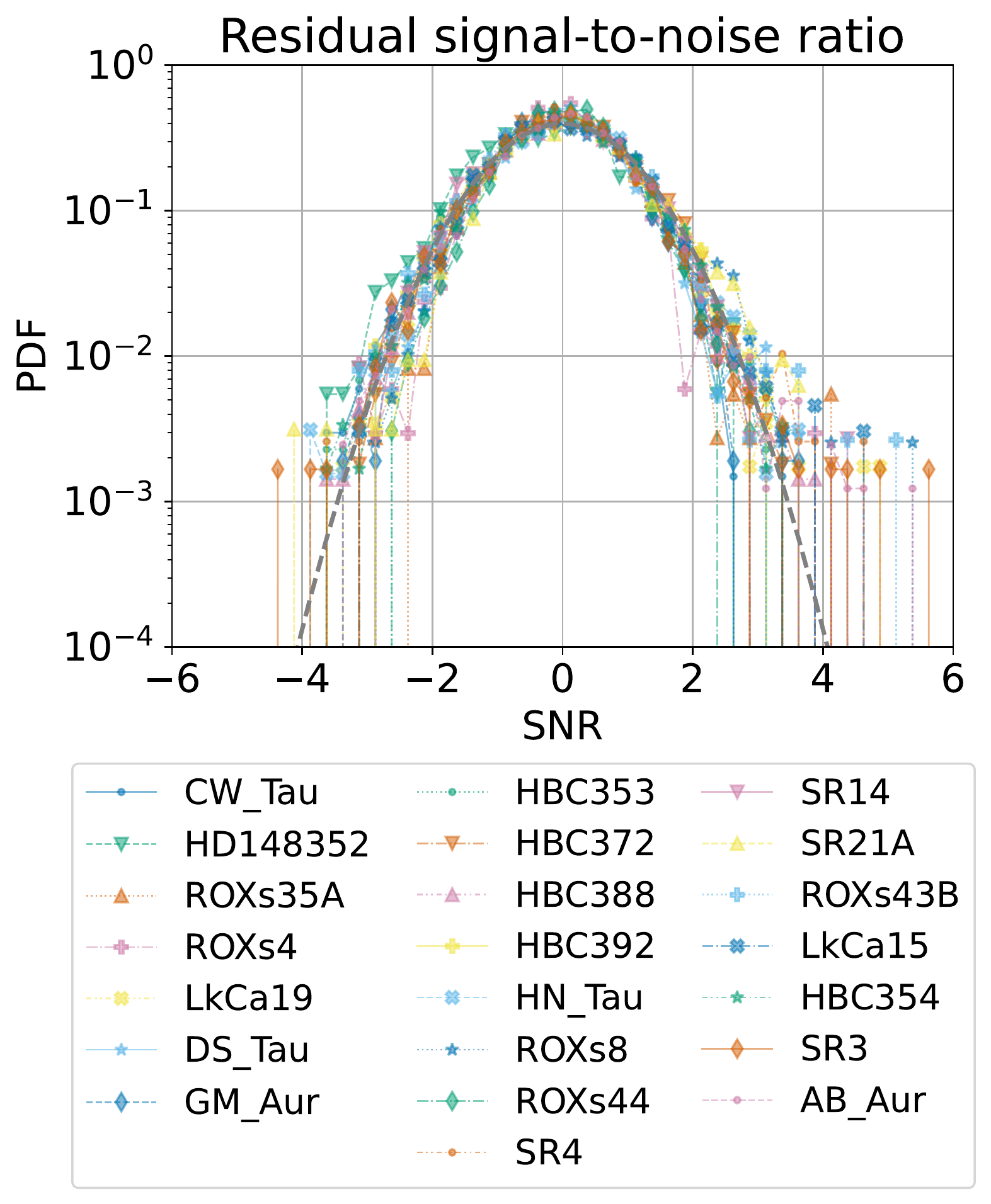}{0.49\textwidth}{(a)}
		\fig{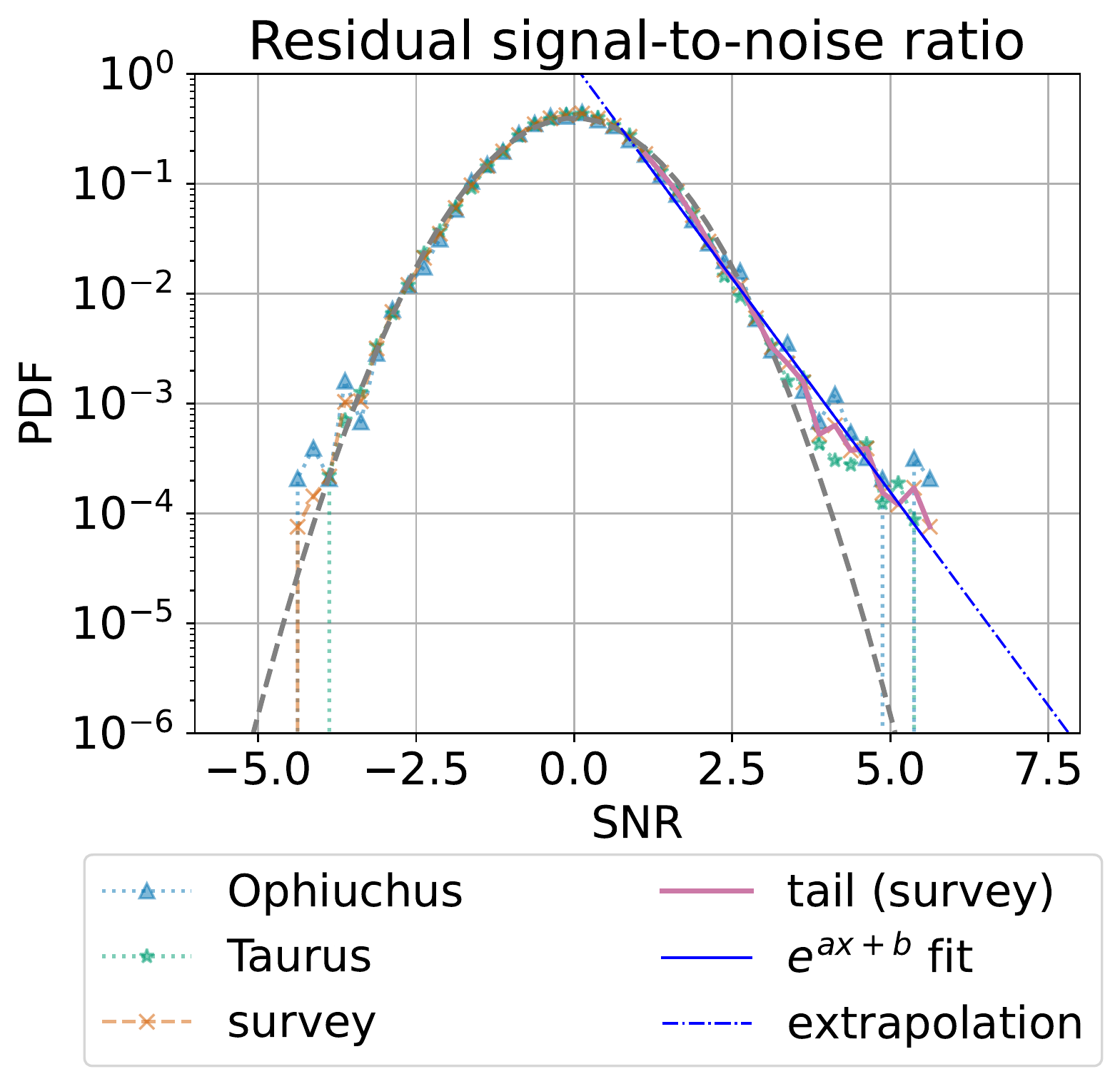}{0.49\textwidth}{(b)}
          }
  \caption{\label{figure:hist_snr} Probability distribution function of residual signal-to-noise values (a) for all (but one) of our targets individually, and (b) for Ophiuchus, Taurus, and all targets. (b) also shows the tail (purple) of the survey histogram that was fitted to an exponential decay (blue) to get extrapolated values at under-sampled S/N's. We do not include the contribution from Em* SR 9, as the values are biased by the bright binary companion, and mask HD 148352 B. Ideal Gaussian behavior is the black dashed bell curve}
\end{figure*}

Using the method described in Section \ref{subsec:planet_search}, we generate S/N maps for each of the 23 targets. These are presented in \reviewedit{Appendix \ref{app:detecmaps}}. 
Figure \ref{figure:hist_snr} plots the histogram of the combined S/N for all the targets in our survey. The tail of the distribution of S/N values does not perfectly follow a Gaussian distribution, which needs to be accounted for when defining the detection threshold. We choose a detection threshold that sets the probability of false positives to match that of a $5\sigma$ threshold when assuming a Gaussian distribution. To do so, we extrapolate from a least-squares fit the tail of the survey histogram to an exponential $\text{PDF} = \exp(a\; \secondreviewedit{\times} \; \text{S/N} + b)$ and derive an equivalent detection threshold of $8\times\sigma_p$.

We observe only two features with S/N $>$ 8: the well-studied binary companion of Em* SR 9 \citep{ghez_multiplicity_1993} and the new detection of an M dwarf candidate around HD 148352, as discussed in Section \ref{subsec:hd148352}. 

\subsection{Contrast curves}

\begin{figure}[htpb]
\centering
\includegraphics[width=\columnwidth]{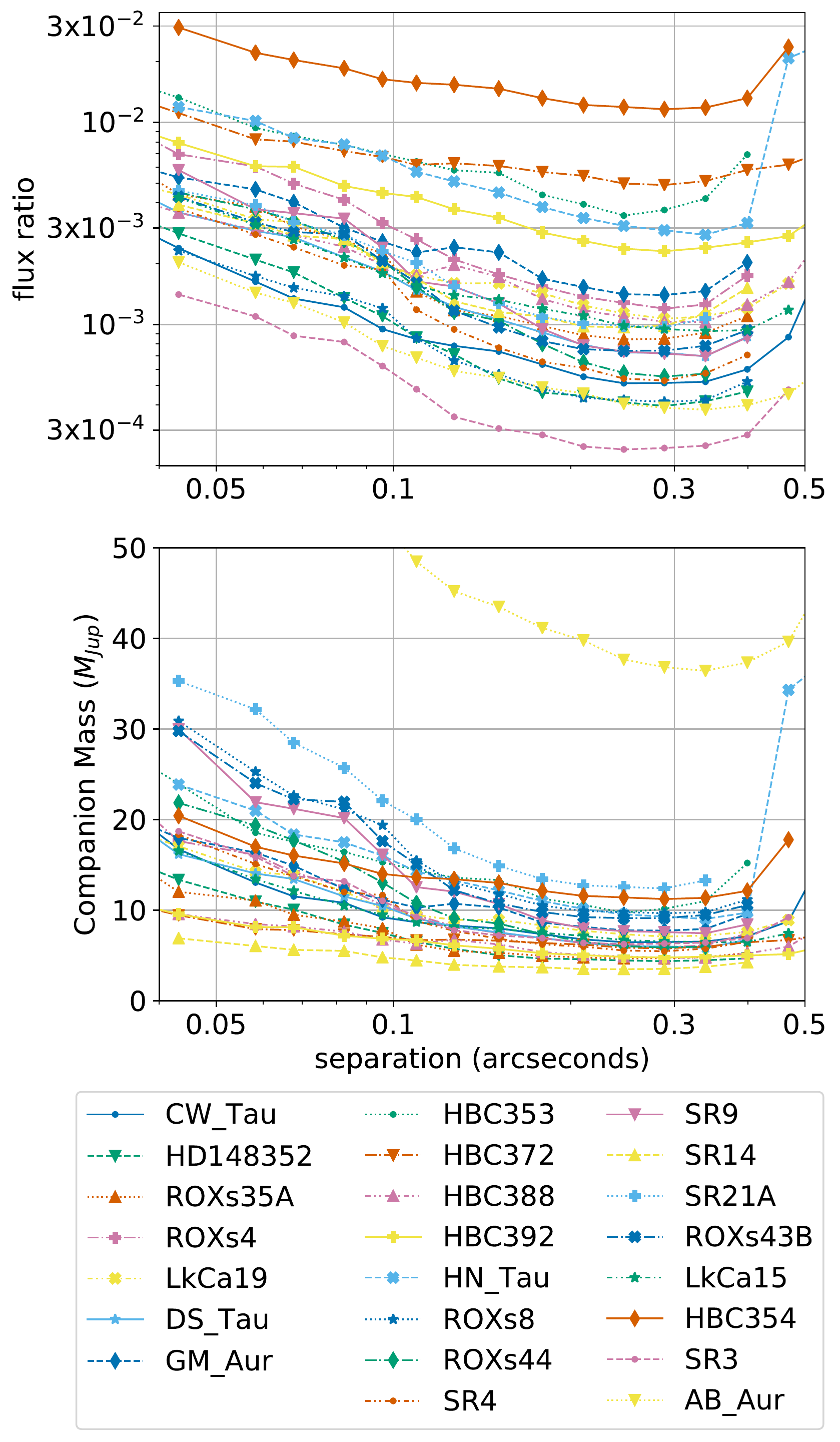}
\caption{\label{figure:combined_contrast} Detection limits for all 23 targets of the survey as a function of angular separation from the star. The sensitivity is shown in terms of the planet-to-star flux ratio (Kn5 filter) in the upper panel, and in terms of the companion mass in the lower panel. The flux to mass conversion was done using a BT-Settl evolutionary grid \citep{allard_bt-settl_2014} assuming an age of 2 Myr.  The detection threshold was chosen to $8\times\sigma_p$ accounting for the non-Gaussianity of the S/N residuals. This threshold matches an equivalent false positive rate of a $5\times\sigma$ threshold assuming a Gaussian distribution.} 
\end{figure}

\begin{figure}[htpb]
\centering
\includegraphics[width=\columnwidth]{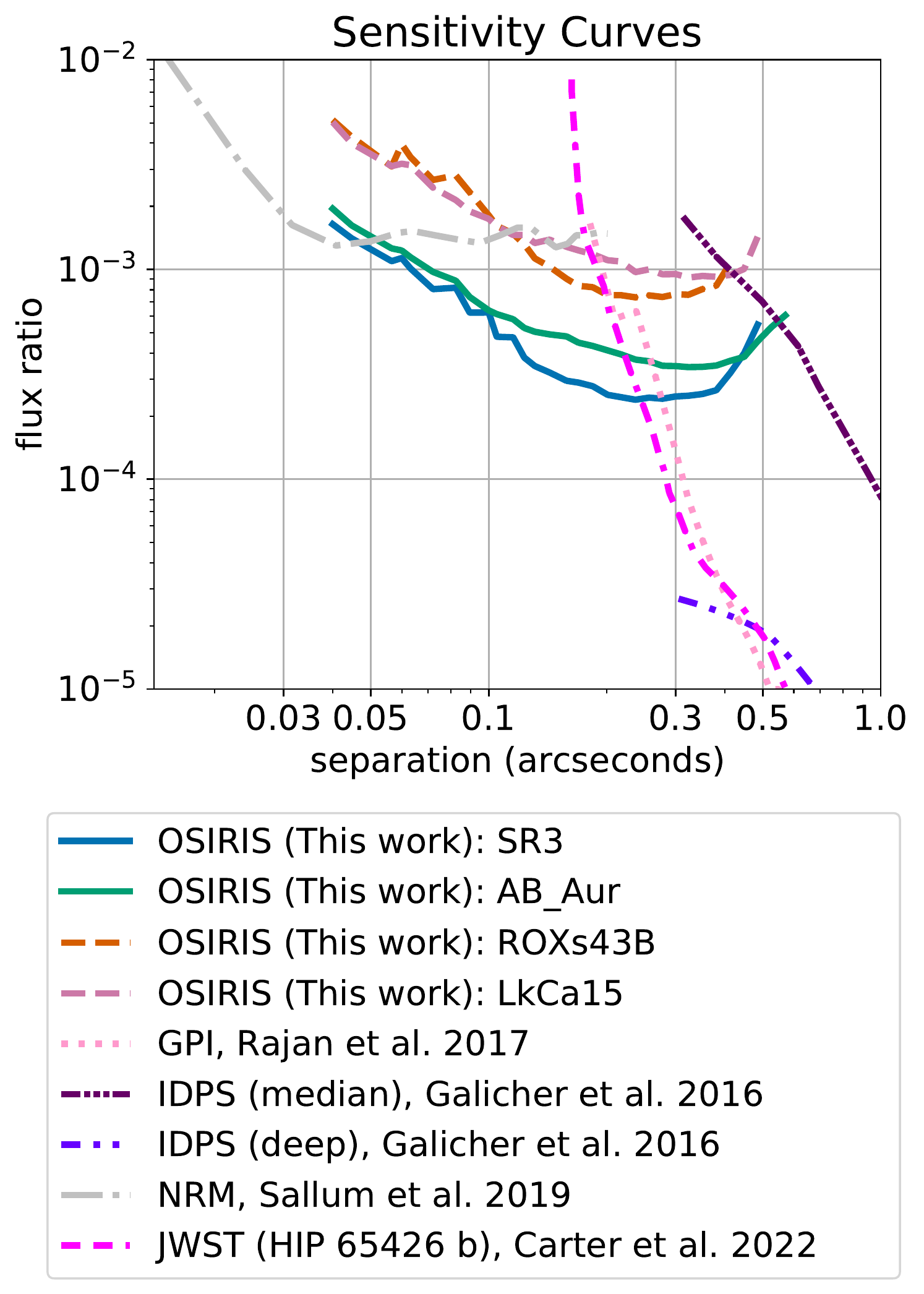}
\caption{\label{figure:compare_contrast} Sensitivity of Keck/OSIRIS at Kn5 (solid/dash) compared with K-band sensitivities of other direct imaging surveys (dot/dash-dot). LkCa 15 (purple) and ROXs 43B (red) illustrate the median sensitivity of the survey, while the best performance was achieved for AB Aurigae (green) and Em* SR 3 (blue).} 
\end{figure}

Figure \ref{figure:combined_contrast} plots the sensitivity achieved around each target in terms of flux ratio and companion mass as a function of angular separation using the $8\times\sigma_p$ detection threshold derived in Section \ref{subsec:snr_maps}. Assuming an age of 2 Myr, OSIRIS is therefore sensitive to companions at the boundary between planets and brown dwarfs at 100 mas in Ophiuchus and Taurus. 
The final sensitivity varies as a function of stellar magnitude, spectral type, observing conditions, and exposure time. 

Figure \ref{figure:compare_contrast} compares our survey's best and median sensitivities to scaled sensitivities of contemporary direct imaging surveys in the K band. These nominal examples include the Gemini Planet Imager (GPI; \citet{macintosh_first_2014}), which used high contrast imaging, extreme adaptive optics (XAO), diffraction control, a coronagraph, and a low-resolution spectrograph ($R \approx 30 - 70$); we scale GPI sensitivities from \citet{Rajan_2017} to those of thirty-minute exposures, which is roughly our exposure time for each target. \reviewedit{However, this does not account for the reduced sky rotation that comes with a shorter exposure time, which would reduce the GPI sensitivity even more, especially at such small projected separation.} We use \citet{sallum_comparing_2019} for simulated sensitivities for non-redundant masking using a pupil-plane mask. We also plot the deepest sensitivities of the International Deep Planet Survey \citep[IDPS;][]{Galicher_idps_2016} that uses a composite of four different instruments with adaptive optics systems. Similar to Keck/OSIRIS, these instruments do not use XAO. Finally, we include measured sensitivities for JWST/NIRCam (Near-Infrared Camera) from Early Release Science $\sim 40$-minute coronagraphic observations of HIP 65426 b \citep{jwst_ers} in their F250M filter (mean wavelength $2.523$ $\mu$m, bandwidth $0.179$ $\mu$m). At separations of less than 300 milliarcseconds, which is closer to the star than allowed by typical coronagraphs, we estimate better sensitivities than these counterparts.

\subsection{\label{subsec:depth_search} \reviewedit{Mid-Survey Depth of Search}}

\begin{figure}[htpb]
\centering
\includegraphics[width=\columnwidth]{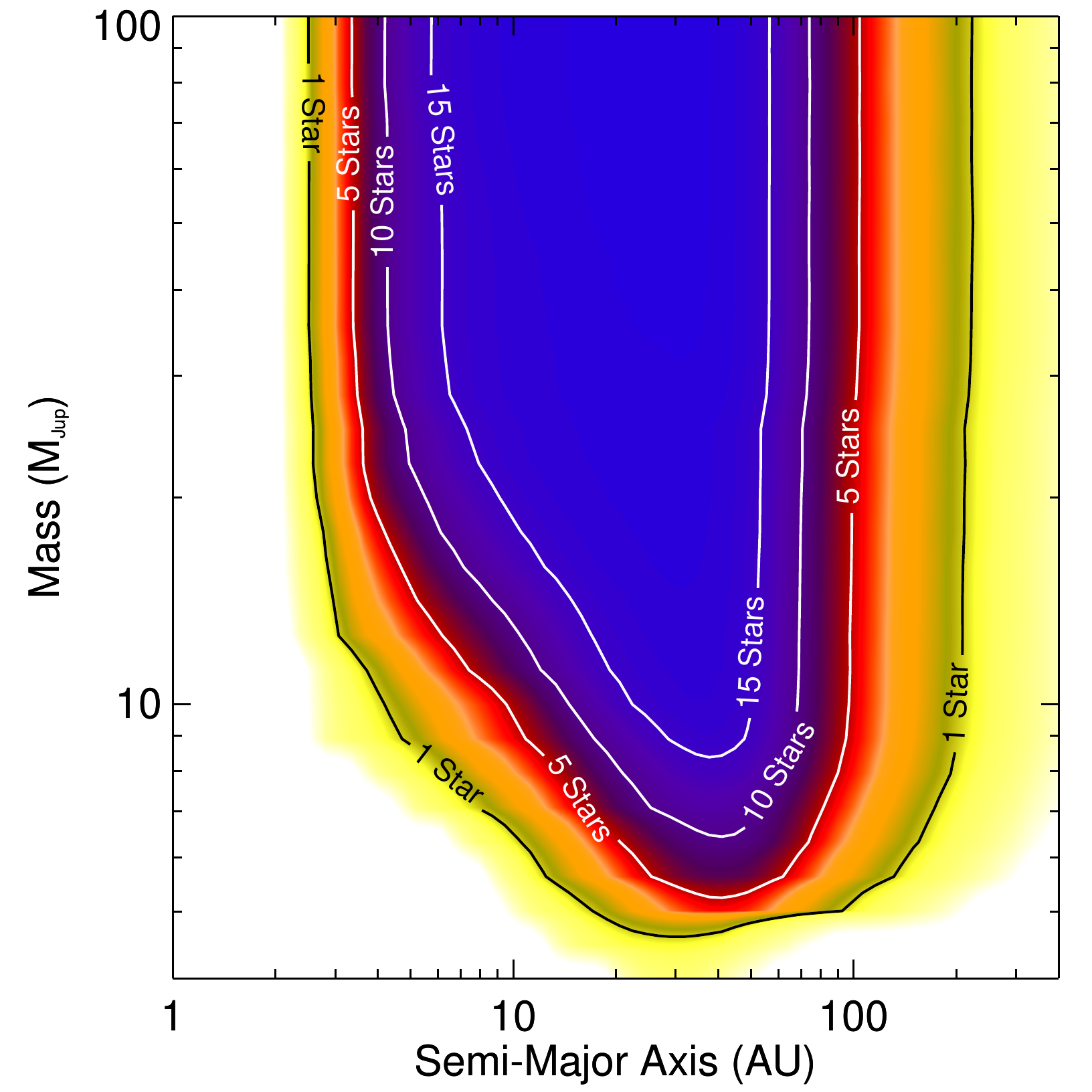}
\caption{\label{figure:depth_search} \reviewedit{Depth of search for the 20 stars observed in this Keck/OSIRIS survey. The contours denote the number of stars to which the survey is complete for planets and brown dwarfs, as a function of separation (in au) and mass (in Jupiter masses). Our survey is probing similar parameter space to extreme adaptive optics surveys of stars in young moving groups \citep{nielsen_gemini_2019}, despite our target stars being about 3 times further away.}} 
\end{figure}

\reviewedit{We present the depth of search for the 20 stars observed in Ophiuchus and Taurus as part of our Keck/OSIRIS survey in Figure \ref{figure:depth_search}. This is calculated using the procedure described in \citet{nielsen_gemini_2019}: the completeness is determined for each target using the MCMC framework developed in \citet{Nielsen2008, Nielsen2010, Nielsen2013}; summing completeness maps for all targets gives a map representing the depth of search of our survey till date. We plan to present and discuss results for the completeness of the full survey in future work.}

\reviewedit{The parameter space probed by this pilot study is very similar to that reached by the first 300 stars of the GPIES campaign \citep{nielsen_gemini_2019}, albeit at a much younger age, with sensitivity to brown dwarfs from $\sim$5-200 au, and giant planets from $\sim$10-100 au.  This underscores our sensitivity to small-separation companions: the median distance to the GPIES sample was 45 pc \citep{nielsen_gemini_2019}, \secondreviewedit{3 times} closer than the stars in our sample.  Given the relatively low occurrence rate of wide-separation companions (\citet{nielsen_gemini_2019} report, for 10-100 au, an occurrence rate for 13-80 M$_{Jup}$ brown dwarfs of 0.8$^{+0.8}_{-0.3}$ \%, and 5-13 M$_{Jup}$ giant planets of 8.9$^{+5.0}_{-3.6}$ \%), a larger number of stars would be required to expect significant detections of these companions, if these occurrence rates from $\sim$100 Myr stars are the same for much younger systems.  Thus, our technique represents a unique opportunity to probe brown dwarf and exoplanet demographics as a function of age, by probing a similar mass and semi-major axis range in star-forming regions more distant than current surveys of nearby moving groups.}

\section{Discussion}
\label{sec:discussion}

Our results demonstrate how integral field spectrographs with moderate to high spectral resolution can be used to search for planets \reviewedit{at small angular separations}. Typically, high spectral resolution instruments can be used for characterization only after we have precise values for the location of the companion; for example, the Keck Planet Imager and Characterizer \citep{lens.org/038-174-150-528-679, lens.org/039-865-813-267-30X, lens.org/160-034-496-708-501} uses fibers and collects data from a single \reviewedit{position on} the sky. On the other hand, integral field spectrographs such as Keck/OSIRIS allow us to use their moderate spectral resolution to search for previously undetected companions, \reviewedit{over their entire FOV}.  A caveat of using the information included in molecular lines is that the method will be less sensitive to embedded planets, where spectral features can be muted \citep{Cugno2021}. Our sensitivity will depend on the \reviewedit{differing} spectral types of the planet and the star, whereas NRM is independent of the stellar spectral type for example.

Keck/OSIRIS was not specifically designed for high-contrast imaging, but this work shows that it could detect planets at smaller separations than coronagraphic instruments and at higher contrast than aperture masking techniques. This is consistent with similar work studying the prospect of future moderate spectral resolution IFSs such as JWST/MIRI \citep{Patapis2022}, JWST/NIRSpec \citep{Llop-Sayson2021}, ELT/HARMONI \citep{Houlle2021}, or ELT/METIS \citep{Carlomagno2020}. Existing studies have been based on simulations or single targets \citep[e.g.,][]{Hoeijmakers2018}, but this work is the first planet search survey. By including a sample of \reviewedit{20} stars, we can study the statistics of the final S/N maps down to lower false positive rates (see Figure \ref{figure:hist_snr}) and highlight some remaining limitations of the method that are difficult to identify in single datasets. The non-gaussianity of the tail of the S/N maps histograms and the need for additional components to the forward model (e.g., PCA components, telluric-only model) show that further progress in the data modeling is still possible and also needed. \reviewedit{It is also possible to model correlated noise with a non-diagonal covariance in the statistical framework used in \breadsnospace, but this is left for future work.} Improvements to the instrument calibration, such as a variable line spread function, and direct forward modeling of the detector images to avoid interpolations in the OSIRIS DRP could be a path forward to improving these systematics. Additionally, using OSIRIS's broader K-band filter could allow us to improve our sensitivity, as we will cover a larger spectral range with each spaxel's data while compromising to a smaller field of view around the star. This survey is also a pilot program aimed at preparing the future science campaigns of the KAPA adaptive optics upgrades on Keck I \citep{Lu2020, Wizinowich2022}. 

Additionally, the forward modeling of the continuum presented in this work represents an alternative to high-pass filtering and continuum normalization. It can retain more information about the spectrum of the companion and could improve the characterization of planetary atmospheres.

Recent advances with \reviewedit{extreme adaptive optics (XAO)} instruments have resulted in significant sensitivity improvements at sub-arcsecond separations. 
Our technique could prove highly relevant for a future high-resolution integral field spectrograph with XAO.
This type of analysis that leverages the high-resolution spectral signatures of planetary atmospheres can also be combined with coronagraphy and other well-established data reduction techniques such as \reviewedit{reference-star differential imaging} \citep[RDI; e.g.,][]{Xuan2018, Wahhaj2021}.

\section{Conclusion}
\label{sec:conclusion}

We have explored the detection of planets at small projected separations from their host stars by searching for their distinct spectral features hidden amongst diffracted starlight in data from moderate-resolution integral field spectrographs. We analyzed the data of a \reviewedit{20}-star pathfinder survey with Keck/OSIRIS ($R\approx4,000$) of two nearby star-forming regions Ophiuchus and Taurus. We demonstrate an improved sensitivity under 300 milliarcseconds compared to other detection techniques, enabling the detection and study of larger populations of gas giant planets. Prospects also include extending the applicability of our forward model approach, through the framework of \breadsnospace, to the next generation of instruments such as NIRSpec, the near-infrared integral field spectrograph aboard JWST, as well as similar instruments on the future ELTs.

\section*{Acknowledgments}
This research was funded in part by the Gordon and Betty Moore Foundation through grant GBMF8550 and by NASA ROSES XRP award 80NSSC19K0294 to M.L. and D.M.. J.-B. R. acknowledges support from the David and Ellen Lee Prize Postdoctoral Fellowship. S.A.'s work was supported by the Rita A. and Øistein Skjellum SURF Fellowship.

K.K.W.H., Q.M.K., and T.S.B. acknowledge support by the National Aeronautics and Space Administration under ROSES Grant No. 80NSSC21K0573 issued through the Astrophysics Division of the Science Mission Directorate.

The W. M. Keck Observatory is operated as a scientific partnership among the California Institute of Technology, the University of California, and NASA. The Keck Observatory was made possible by the generous financial support of the W. M. Keck Foundation. We also wish to recognize the very important cultural role and reverence that the summit of Mauna Kea has always had within the indigenous Hawaiian community. We are most fortunate to have the opportunity to conduct observations from this mountain.

\vspace{5mm}
\facilities{KeckI (OSIRIS)}

\software{
astropy\footnote{\url{http://www.astropy.org}} \citep{astropy:2013, astropy:2018, astropy:2022}, 
Matplotlib\footnote{\url{https://matplotlib.org}} \citep{Hunter:2007},
breads\footnote{\url{https://github.com/jruffio/breads}}\footnote{\url{https://github.com/shubhagrawal30/using-breads}} (this work; \citet{agrawal_direct_2022}),
emcee\footnote{\url{https://github.com/dfm/emcee}}\citep{emcee},
\texttt{corner} \citep{corner}
}

\clearpage
\appendix

\FloatBarrier
\section{Combined Detection Maps}
\label{app:detecmaps}

\begin{figure}[!htpb]
\centering
\includegraphics[width=0.9\textwidth]{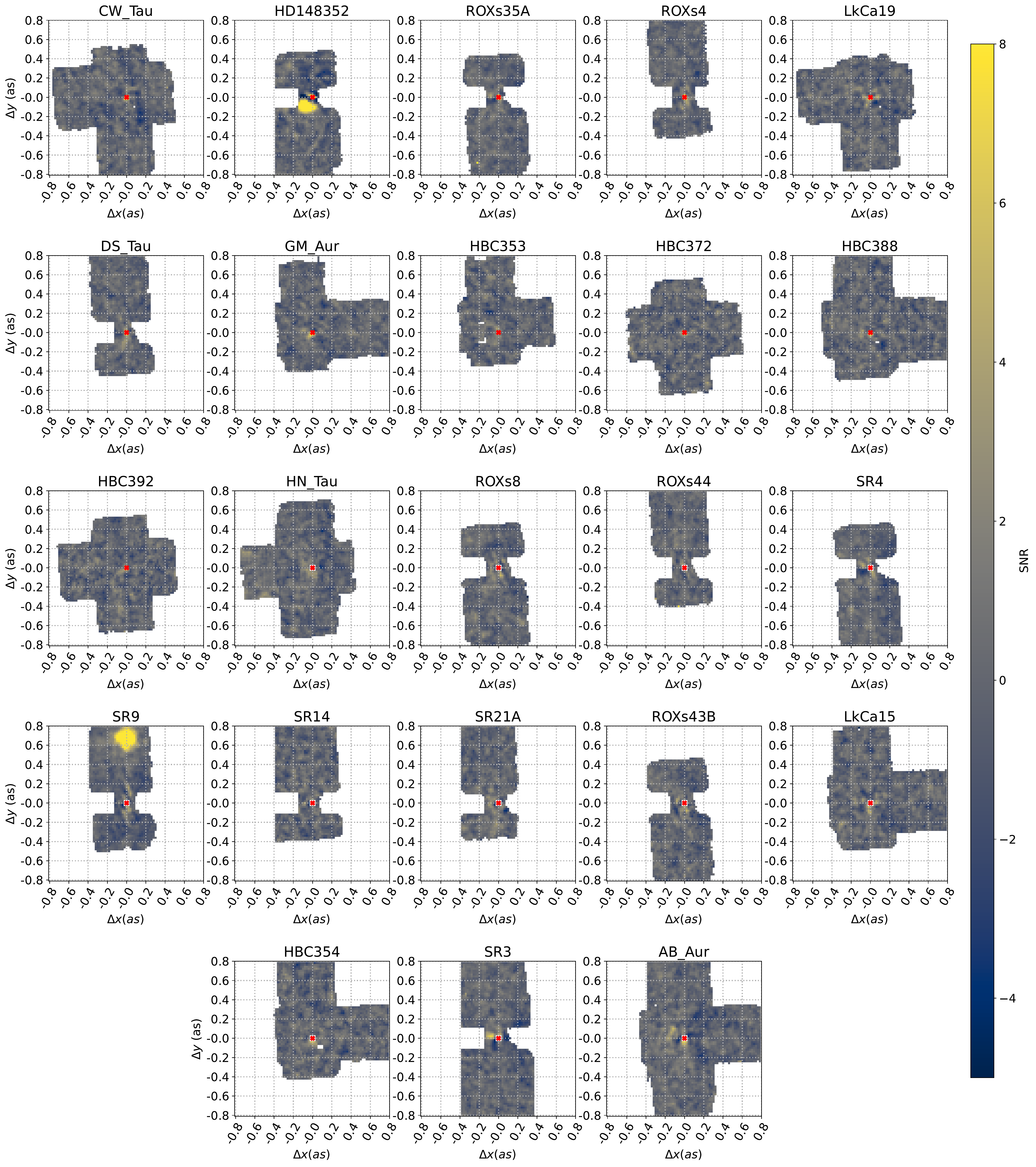}
\caption{\label{figure:SNR_maps} \reviewedit{Signal-to-noise spatial detection maps for our pathfinder survey with Keck/OSIRIS, obtained after combining all data for each target at this mid-survey stage. The plate scale is 0.02" per spaxel.}} 
\end{figure}
\FloatBarrier

\section{\reviewedit{Table of Observations}}

\begin{table*}[h]
\centering
\begin{tabular}{|c||c|c|c|c|c|c|c|c|}
\hline
Star & RA & Dec & Type & $M/M_\odot$ & $R_{mag}$ & $K_{mag}$ & \reviewedit{$p$} & Shorthand\\
\hline
\multicolumn{9}{|c|}{Ophiuchus Target List} \\
\hline
Em* SR 3 & 16:26:09.31 & -24$ ^{\circ}$34'12.1'' & B6 & 5.20 & 10.6 & 6.504 & 7.1448 & SR3\\
Em* SR 21A & 16:27:10.27 & -24$ ^{\circ}$19'12.7'' & G1 & 3.69 & 13.25 & 6.719 & 7.3298 & SR21A\\
Em* SR 4 &16:25:56.15 & -24$ ^{\circ}$20'48.1'' & K4.5& 1.35&12.11 & 7.518 & 7.4208 & SR4\\
ROXs 44  & 16:31:33.46 & -24$ ^{\circ}$27'37.3'' & K3 &    2.07 & 12.35 & 7.61 & 6.8343 & ROXs44\\
ROXs 8 & 16:26:03.02 & -24$ ^{\circ}$23'36.0'' & K1& 2.95 & 9.34 & 6.227 & 7.3439 & ROXs8\\
ROXs 4  & 16:25:50.52 & -24$ ^{\circ}$39'14.5'' & K5.5 & 1.15& 13.88& 8.33 & 7.1771 & ROXs4\\
ROXs 35A & 16:29:33.97 & -24$ ^{\circ}$55'30.3'' & K3&    2.07& 12.41& 8.531 & 6.9527 & ROXs35A\\
ROXs 43B & 16:31:20.19 & -24$ ^{\circ}$30'00.9'' &K5 &1.20& 12* & 7.089 & 6.9199 & ROXs43B\\
Em* SR 14 &16:29:34.41 & -24$ ^{\circ}$52'29.2'' & G4 &    3.64& 10.04 & 8.878 & 8.907 & SR14\\
Em* SR 9 & 16:27:40.28 & -24$ ^{\circ}$22'04.0'' & K5 &    1.20& 11.87&  7.207 & 7.5792 & SR9\\
\hline
\multicolumn{9}{|c|}{\reviewedit{Non-Member Targets}} \\
\hline
\reviewedit{HD 148352} & 16:28:25.16 & -24$^{\circ}$45'00.9'' & F2 & 1.52 & 7.3 & 6.511 & 13.0833 & HD148352\\
\reviewedit{HBC 353} & 03:54:30.18&+32$ ^{\circ}$03'04.4'' & G5 & 3.42 &9.862& 3.4188 & 3.958 & HBC353\\
\reviewedit{HBC 354} & 03:54:35.56&+25$ ^{\circ}$37'11.2''& K3 & 1.40 & 13.79& 11.095 & 3.6716 & HBC354\\
\hline
\multicolumn{9}{|c|}{Taurus Target List} \\
\hline
AB Aurigae & 04:55:45.85 & +30$ ^{\circ}$33'04.3'' & A0Ve & 4.00 & 7.05 & 4.23 & 6.4127 & AB\_Aur\\
CW Tauri & 04:14:17.00 & +28$ ^{\circ}$10'57.8'' & K0Ve & 1.40 & 12.36 & 7.127 & 7.6017 & CW\_Tau\\
DS Tauri & 04:47:48.60 & +29$ ^{\circ}$25'11.2'' & K4Ve & 0.97 & 12.3 & 8.036 & 6.315 & DS\_Tau\\
LkCa 15 &04:39:17.79&+22$ ^{\circ}$21'03.4''& K5Ve & 0.97 & 12.03 & 8.163 & 6.3619 & LkCa15\\
LkCa 19 & 04:55:36.97&+30$ ^{\circ}$17'55.1''& K0Ve & 2.42 &11.12&8.148 & 6.3598 & LkCa19\\
HBC 388 & 04:27:10.57&+17$ ^{\circ}$50'42.7'' & K1e & 2.10 & 10.22& 8.296 & 8.4115 & HBC388\\
GM Aurigae & 04:55:10.98&+30$ ^{\circ}$21'59.4''& K3Ve & 1.40 & 13.1 & 8.283 & 6.3248 & GM\_Aur\\
HN Tauri & 04:33:39.36&+17$ ^{\circ}$51'52.3'' & K5e & 0.97 & 13.4 & 8.384 & 7.4391 & HN\_Tau\\
HBC 392 & 04:31:27.18&+17$ ^{\circ}$06'24.8'' & K5e & 0.97 &12.1& 9.497 & 7.2018 & HBC392\\
HBC 372 & 04:18:21.48&+16$ ^{\circ}$58'47.0'' & K5 & 0.97 &13.26&10.464 & 5.5266 & HBC372\\
\hline
\end{tabular}
\caption{\label{table:targets} Our 23 targets in Ophiuchus and Taurus star-forming regions. \textit{RA} and \textit{Dec} list right ascension (in hrs:mins:secs) and declination in (degrees, arcmins, arcsecs) respectively. \textit{Type}, $M/M_\odot$, $R_{mag}$, $K_{mag}$, and \reviewedit{$p$} refer to spectral type, stellar mass in units of solar mass, R-band magnitude (relevant for NGS), K-band magnitude (relevant for observations), and parallax in milli-arcseconds. \textit{Shorthand} lists the abbreviation used in this work to denote this specific target. Values are from \citet{cheetham_mapping_2015} for Ophiuchus and from \citet{kenyon_pre-main-sequence_1995, esa_hipparcos_1997} for Taurus targets, \reviewedit{as well as from CDS/SIMBAD}. HD 148352 is not in the Ophiuchus cluster, but rather is an interloper, as discussed in Section \ref{subsec:hd148352}. \reviewedit{HBC 353 and HBC 354 are plausible non-members, misclassified as Taurus members in \citet{kenyon_pre-main-sequence_1995}.}}
\end{table*}

\begin{table*}[t]
\centering
\begin{tabular}{|c||c|c|c|c|c|}
\hline
\multicolumn{6}{|c|}{Ophiuchus Science Observations} \\
\hline
Star & Date & Dithers & Frames & Exposure Time (s) & Total Time (min)\\
\hline
Em* SR 3 & 2021/26/06 & 1, 3, 1 & 11, 21, 6 & 30 & 57.67\\
 & 2021/26/07 & 4, 2 & 6, 4 & 20 &\\
 & 2021/26/08 & 3, 4 & 3, 3 & 20 &\\
Em* SR 21A& 2021/26/07 & 2 & 11 & 30 & 11\\
Em* SR 4& 2021/26/07 & 4 & 5 & 90 & 30\\
ROXs 44& 2021/26/07 & 4 & 5 & 90 & 30\\
ROXs 8& 2021/26/07 & 4 & 11 & 20 & 29.33\\
ROXs 4& 2021/26/07 & 5 & 4 & 120 & 40\\
ROXs 35A& 2021/26/08 & 5 & 4 & 120 & 40\\
Em* SR 14& 2021/26/08 & 4 & 4 & 120 & 32\\
ROXs 43B& 2021/26/08 & 4 & 5 & 90 & 30\\
Em* SR 9& 2021/26/08 & 4 & 4 & 90 & 24\\
\hline
\multicolumn{6}{|c|}{\reviewedit{Non-Member Targets}} \\
\hline
\reviewedit{HD 148352} & 2021/26/06 & 3 & 21 & 30 & 31.5\\
\reviewedit{HBC 353} & 2021/10/20 & 3, \textbf{\textit{1}}* & 1 & 300 & 20\\
\reviewedit{HBC 354} & 2021/10/20 & 3, \textbf{\textit{2}}* & 2 & 300 & 50\\
\hline
\multicolumn{6}{|c|}{Taurus Science Observations} \\
\hline
AB Aurigae & 2021/10/18 & 2, \textbf{\textit{2}}*, 3, \textbf{\textit{4}}* &10 & 10, 10, 4, 4 & 17.33\\
& 2021/10/19 & 3, \textbf{\textit{3}}*, \textbf{\textit{3}}* & 6 & 4 &\\
& 2021/10/20 & 3, \textbf{\textit{3}}* & 6 & 4 &\\
CW Tauri & 2021/10/18 & 3, \textbf{\textit{3}}* & 10 & 30 & 30\\
DS Tauri & 2021/10/18 & 3, \textbf{\textit{2}}*& 7 & 90 & 52.5\\
LkCa 15 & 2021/10/18 & 3, \textbf{\textit{3}}* & 5 & 90 & 45\\
LkCa 19 & 2021/10/19 & 3, \textbf{\textit{3}}* & 4 & 90 & 36\\
HBC 388 & 2021/10/19 & \textbf{\textit{3}}*, 2& 4 & 90 & 30 \\
GM Aurigae & 2021/10/19 & \textbf{\textit{3}}*, 2 & 4 & 90 & 30\\
HN Tauri & 2021/10/19 & 3, \textbf{\textit{1}}* & 4, 2 & 90 & 21\\
HBC 392 & 2021/10/20 & \textbf{\textit{3}}*, 3 & 1 & 300 & 30\\
HBC 372 & 2021/10/20 & \textbf{\textit{3}}*, 3 & 1 & 300 & 30\\
\hline
\end{tabular}
\caption{\label{table:science} Science observations of the 23 targets. We include the date, number of dithers per target, frames per dither position, exposure time (in seconds) per frame, and total exposure time (in minutes) over the full survey. For Taurus targets \reviewedit{(as well as HBC 353 and HBC 354)}, the dither positions that are \textbf{in bold}, \textit{italicized}, \reviewedit{and marked with an asterisk (*)} were taken after the field of view was rotated by ninety degrees (to counter pixel bleeding as described in Section \ref{subsec:forward_model}). \reviewedit{The total observation time (in the rightmost column) is given once per target and represents the aggregate of all the science observations taken for the target in this survey to generate the results of this paper.} \\ \reviewedit{Multiple values (separated by commas) in the \textit{Dithers} column represent multiple independent sequences of images taken with dithering in between. For these rows, when the \textit{Frames} column has only one value, the same number of frames were taken for each dither position; in case, different numbers of frames were taken for these dither positions (listed in the same row), the respective values of frames-per-dither are listed in the \textit{Frames} column.}}
\end{table*}

\begin{table*}[t]
\begin{tabular}{|c||c||c|c|}
\hline
\multicolumn{4}{|c|}{Observing Conditions and Calibration Sky Observations} \\
\hline
Date & DIMM Seeing ('') & Frames & Exposure (s) \\
\hline
2021/26/06 & $0.59\pm0.16$ & 2 & 600 \\
2021/26/07 & $0.66\pm0.17$  & 1 & 600 \\
2021/26/08 & $0.53\pm0.10$  & 2 & 600 \\
\hline
2021/10/18 & $0.54\pm0.17$  & 2 & 600 \\
2021/10/19 & $0.60\pm0.28$  & 2 & 600 \\
2021/10/20 & $0.62\pm0.13$  & 2 & 300 \\
\hline
\end{tabular}
\caption{\label{table:skycalib} Atmospheric seeing measured by Maunakea-DIMM for each observing night, and the long exposure sky images taken to perform wavelength and resolution calibration using OH emission lines, as described in Section \ref{sec:observations}}
\end{table*}

\begin{table*}[t]
\centering
\begin{tabular}{|c||c|c|c|c|c|}
\hline
\multicolumn{6}{|c|}{Ophiuchus Calibration Observations} \\
\hline
Star & Date & Sequences & Frames & Exposure (s) & Targets\\
\hline
HIP 73049 & 2021/26/06 & 6 & 1 & 2 & SR3\\
& 2021/26/07 & 7 & 1 & 1.5 & SR4\\
& 2021/26/08 & 7 & 1 & 1.5 & ROXs35A \\
Em* SR 3 & 2021/26/06 & 1, 3 & 11, 21 & 30 & HD148352, SR3, SR21A\\
& 2021/26/06 & 1 & 6 & 30 & -\\
 & 2021/26/07 & 4, 2 & 6, 4 & 20 & ROXs44, ROXs8\\
 & 2021/26/07 & 2 & 4 & 20 & ROXs4 \\
 & 2021/26/08 & 3, 4 & 3, 3 & 20 & SR14, ROXs43B\\
 & 2021/26/08 & 4 & 3 & 20 & SR9\\
\hline
\multicolumn{6}{|c|}{Taurus Calibration Observations} \\
\hline
AB Aurigae & 2021/10/18 & 2, \textbf{\textit{2}}* &10 & 10, 10 & AB\_Aur, CW\_Tau\\
 & 2021/10/18 & 3 &10 & 4 & DS\_Tau\\
 & 2021/10/18 & \textbf{\textit{4}}* &10 & 4 & LkCa15\\
& 2021/10/19 & 3 & 6 & 4 & LkCa19\\
& 2021/10/19 & \textbf{\textit{3}}* & 6 & 4 & HBC388, GM\_Aur\\
& 2021/10/19 & \textbf{\textit{3}}* & 6 & 4 & HN\_Tau\\
& 2021/10/20 & 3 & 6 & 4 & HBC354\\
& 2021/10/20 & \textbf{\textit{3}}* & 6 & 4 & HBC392, HBC372, HBC353\\
\hline
\end{tabular}
\caption{\label{table:calibration} Observations of A0 standard stars \reviewedit{(listed in the leftmost column)}, taken to calculate sky transmission and perform telluric calibration, as described in Section \ref{sec:observations}. \reviewedit{We also list the targets calibrated using each set of observations in the rightmost column.} We include the date of observations, number of sequences, frames for each sequence, and exposure time per frame in seconds. Numbers in \textbf{bold}, \textit{italics}, and marked with an asterisk (*) represent that the sequence was taken after rotating the field of view by ninety degrees. Several of these observations also serve as the science data for the corresponding \reviewedit{A0 star}. \\
\reviewedit{Multiple values (separated by commas) in the \textit{Dithers} column represent multiple independent sequences of images taken with dithering in between. For these rows, when the \textit{Frames} column has only one value, the same number of frames were taken for each dither position; in case, different numbers of frames were taken for these dither positions (listed in the same row), the respective values of frames-per-dither are listed in the \textit{Frames} column.}}
\end{table*}

\bibliography{bibliography}{}
\bibliographystyle{aasjournal}

\end{document}